\documentclass[12pt]{article}

\usepackage{latexsym}

\usepackage[font=small,labelfont=bf]{caption}

\usepackage{graphics}
\usepackage{graphicx}
\usepackage{epstopdf}
\usepackage{amssymb}
\usepackage{tabularx}
\usepackage{caption}
\usepackage{subcaption}
\usepackage{dcolumn}
\usepackage{bm}
\usepackage{hyperref}
\usepackage{tabularx}
\usepackage{slashbox}
\usepackage{tikz}
\usetikzlibrary{matrix}
\newsavebox{\tempbox}
\usepackage{gensymb}
\newcommand{\be}{\begin{equation}}
\newcommand{\ee}{\end{equation}}
\newcommand{\ba}{\begin{eqnarray}}
\newcommand{\ea}{\end{eqnarray}}
\newcommand{\ban}{\begin{eqnarray*}}
\newcommand{\ean}{\end{eqnarray*}}

\graphicspath{{./Figures/}}
\begin{document}


\title {Quasi-periodic oscillations from the accretion disk around distorted black holes}

\author{
Efthimia Deligianni$^{1}$\footnote{E-mail: \texttt{efthimia.deligianni@uni-oldenburg.de}}, \, Jutta Kunz$^{1}$\footnote{E-mail: \texttt{jutta.kunz@uni-oldenburg.de}}, \, Petya Nedkova$^{1,2}$\footnote{E-mail: \texttt{pnedkova@phys.uni-sofia.bg}}\\ \\
    {\footnotesize${}^{1}$  Institut f\"{u}r Physik, Universit\"{a}t Oldenburg}\\
  {\footnotesize D-26111 Oldenburg, Germany}\\
   {\footnotesize${}^{2}$ Faculty of Physics, Sofia University,}\\
  {\footnotesize    5 James Bourchier Boulevard, Sofia~1164, Bulgaria }}
\date{}
\maketitle

\begin{abstract}
We study the quasi-periodic oscillations from the accretion disk around the distorted Schwarzschild black hole in the framework of the resonant models. We confine ourselves to the case of a quadrupole distortion which can be caused for example by the accreting matter flow in the vicinity of the compact object. For the purpose we examine the linear stability of the circular geodesic orbits in the equatorial plane and derive analytical expressions for the radial and vertical epicyclic frequencies. We investigate their properties in comparison with the isolated Schwarzschild black hole. Due to the influence of the external matter the vertical epicyclic frequency is not always positive anymore, and the stability of the circular orbits is determined by the interplay between both of the frequencies. As a result, the stable circular orbits do not extend to infinity, but are confined to a finite annular region between an inner and an outer marginally stable orbit. In addition, the degeneracy between the vertical epicyclic and the orbital frequency, which is characteristic for the Schwarzschild solution, is broken, and there are regions in the parametric space where the radial epicyclic frequency is larger than the vertical one. All these properties allow for much more diverse types of non-linear resonances to be excited than for the isolated Schwarzschild black hole, which can provide an explanation for the observed $3:2$ ratio between the twin-peak frequencies of the quasi-periodic oscillations from the accretion disk.
\end{abstract}

\section{Introduction}
With the recent development of the experimental resources, testing general relativity in the strong field regime became an important line of research. Combining data from electromagnetic and gravitational wave experiments, we aim to put more stringent constraints on the viable gravitational theories and differentiate more precisely between different compact objects, including more exotic self-gravitating systems, like wormholes or naked singularities.

A promising source of information about the gravitational field in the regime of strong interaction is the electromagnetic radiation emitted by the gas in the accretion disks around the compact objects. The imprints of different physical phenomena can be extracted from its X-ray spectrum. Therefore,  a number of experimental missions with constantly improving resolution are being developed for its measurement, such as the satellites LOFT \cite{Feroci:2016}, eXTP \cite{Zhang:2016} and
STROBE-X \cite{Wilson-Hodge:2017}.

One of the most intriguing features of the accretion flow are the high-frequency quasi-periodic oscillations (QPOs) observed in the X-ray flux from neutron star binaries, stellar-mass black holes, and a few supermassive active galactic nuclei. They represent a couple of peaks in the X-ray spectrum with frequencies obeying a constant $3:2$ ratio. The twin peak QPOs scale inversely with the compact object's mass, and depend very weakly on the observed X-ray flux. This suggests that they are not caused by kinematic effects in the accretion disk, but are rather an intrinsic property of the background spacetime carrying fundamental information about its nature.

The precise physical mechanism of the formation of the QPOs is currently unknown. However, the stable $3:2$ scaling between the two frequencies strongly suggests that they can originate from some resonant process taking place in the accretion disk's oscillations. Resonant models describing the QPOs' formation were developed within the thin disk approximation. They suggest that due to the physical processes in the accretion disk the particles don't move on perfectly circular orbits in the equatorial plane, but perform small oscillations in the radial and vertical direction around the orbit. In the linear approximation the oscillations are described by two decoupled harmonic oscillators with frequencies called radial and vertical epicyclic frequencies. If we consider further non-linear corrections, they lead to coupling between the two epicyclic frequencies and the excitation of different resonances. The observed twin peak frequencies can be identified either with the epicyclic frequencies or with a linear combination of them, depending on the type of resonance, which is realized, in order to achieve the experimental $3:2$ ratio. Another possibility is that the resonances occur as a result of the coupling between one of the epicyclic frequencies and the orbital frequency on the circular orbit.

The possibility of formation of non-linear resonances due to coupling between the epicyclic frequencies in black hole spacetimes was first investigated in \cite{Aliev:1981}, \cite{Aliev:1986}, where the Kerr(-Newman) black hole was considered, and further generalized by adding an external magnetic field. Later, the idea was revisited in the context of modelling the QPOs in the accretion disks of black holes in the series of works \cite{Abramowicz:2001}-\cite{Torok:2005} using again the Kerr solution in order to describe the compact object. Recently, the QPOs were viewed as one of the suitable phenomena for differentiating observationally between gravitational theories and testing the Kerr hypothesis \cite{Stella:1999}. Therefore, the resonant models were applied in different black hole spacetimes, including braneworld black holes \cite{Stuchlik:2009}, \cite{Aliev:2013}, quasi-Kerr metrics describing small deviations from the Kerr solution \cite{Johannsen:2011}, or the Tomimatsu-Sato solution \cite{Stefanov:2013}. Black holes in the Einstein-dilaton-Gauss-Bonnet theory and quadratic gravity were considered in \cite{Macelli:2015}-\cite{Macelli:2017}, while \cite{Bambi:2012} studied the QPOs in the Johansen-Psaltis metric,  which describes a general stationary and axisymmetric spacetime with a regular black hole horizon, extending the Kerr solution in the alternative theories of gravity.

The resonant models depend strongly on the properties of the  epicyclic and orbital frequencies, which on the other hand are determined by the background spacetime. If we denote  the radial and vertical epicyclic frequencies by $\nu_r$ and $\nu_\theta$ and the orbital frequency by $\nu_0$, for the Kerr metric we always have the ordering $\nu_0>\nu_\theta>\nu_r$ for any circular orbit. This restricts the possible resonances that can be excited. For the Schwarzschild black hole the situation is even more degenerate, since the vertical epicyclic frequency and the orbital frequency coincide. Black holes in the modified theories of gravity allow for a greater variety in the behaviour of the characteristic frequencies, and hence for more diverse physical mechanisms for the development of non-linear phenomena.

In this paper we investigate another situation when the resonance profile of the black holes in general relativity is modified. Instead of introducing corrections to the gravitational theory, we consider the influence of the surrounding matter on the black hole spacetime, such as an accretion disk or a binary companion. Usually such back-reaction is omitted, since it is considered to be negligibly small compared to the gravitational field of the compact object. However, when it comes to physical phenomena connected with the geodesic motion, even a small correction can be crucial, since it can change qualitatively the geodesic structure, and lead to observable effects.

In order to describe the interaction between the black hole and the surrounding matter we use a class of exact solutions in general relativity called distorted black holes \cite{Geroch}. They represent local solutions valid in the vicinity of the black hole horizon, which reflect the influence of a general surrounding matter distribution on the compact object's spacetime. The distorted black hole solution is vacuum or electro-vacuum, so it does not contain explicitly the matter sources. It is supposed to be valid up to some physically relevant hypersurface and includes the interior multipole moments expansion in the vicinity of this hypersurface in the form of an infinite sequence of free parameters. A global solution, which is valid in the whole spacetime, is achieved by matching on the boundary hypersurface the interior distorted black hole solution with an exterior solution, which contains the matter sources but no black hole horizon. The particular type of the external matter distribution is specified in the distorted black hole metric by constraining its interior multipolar structure. In the limit when all the interior multipole moments vanish, the solution reduces to an isolated black hole with the same symmetries, e.g. the Schwarzschild or the Kerr black hole.

The idea of distorted black holes was first suggested in the literature in the early work of Doroshkevich et al. \cite{Doroshkevich}. The most general static and axisymmetric distorted black hole solution in general relativity was considered in the classical paper of Geroch and Hartle \cite{Geroch}, where its thermodynamics was also analyzed. Rotating and charged generalizations were further developed and their properties studied \cite{Tomimatsu}-\cite{Abdolrahimi:2015}, including also solutions in higher dimensions \cite{Abdolrahimi:2010}-\cite{Nedkova:2014}. Recently, a series of works investigated the particle and light propagation in the vicinity of distorted black holes. The qualitative behavior of the equatorial geodesics for static black holes with quadrupole distortion was studied in \cite{Shoom:2016}, while \cite{Abdolrahimi:2015b}-\cite{Nedkova:2018} considered the appearance of their local shadow. It was demonstrated that the photon region of the distorted black holes is qualitatively different than that of their isolated counterparts even for an arbitrary small distortion \cite{Nedkova:2018},\cite{Shoom:2017}. This leads to distinct observational effects in their shadow silhouettes, which don't depend on the strength of the interaction with the external matter field but on its mere presence \cite{Nedkova:2018}. When we take into account the influence of the surrounding matter, the shadow image is not just a deformed generalization of the shadow of the corresponding isolated black hole, but we observe in addition the appearance of a series of secondary images.

The purpose of this work is to study the influence of the surrounding matter on the stability of the circular geodesics in the equatorial plane and apply the results for describing the quasi-periodic oscillations from the accretion disk within the resonant models. For simplicity we consider the distorted Schwarzschild solution, since it allows to see clearly the impact of the external matter on the particles' dynamics without interfering with effects from the black hole spin. Since the distorted Schwarzschild black hole should be a static limit of more general rotating distorted black hole solutions, the observed effects should be present also when rotation is added. We further restrict ourselves to the case when the external matter is characterized only by a quadrupole moment. This is the simplest multipolar structure describing matter distributions with an equatorial symmetry, including in particular the case of an accretion disk surrounding the black hole.

The paper is organized as follows. In the next section we present briefly the metric of the distorted Schwarzschild black hole. In section 3 we describe the distribution of the circular geodesics in the equatorial plane, while in section 4 we study their stability with respect to linear perturbations. We derive analytic expressions for the radial and vertical epicyclic frequencies, and investigate their properties. The regions of stability are analyzed for positive and negative quadrupole distortion, as well as the behavior of the epicyclic and orbital frequencies as a function of the radial distance. In section 5 we apply the obtained results to examine the possible resonances, which can be excited as a result of the coupling between some of the characteristic frequencies, and which can explain the observed $3:2$ ratio in the quasi-periodic oscillations. In the last section we discuss our results.

\section{Distorted Schwarzschild black hole}

The distorted Schwarzschild black hole is a static axisymmetric solution to the Einstein equations in vacuum. It is given by the metric \cite{Geroch}, \cite{Breton:1997}
\begin{eqnarray}\label{metric}
ds^{2}&=&-\left(\frac{x-1}{x+1}\right)e^{2U}dt^{2}+m^{2}(x+1)^{2}(1-y^{2})e^{-2U}d\phi^{2}\nonumber\\
&+&m^{2}(x+1)^{2}e^{2(V-U)}\left(\frac{dx^{2}}{x^{2}-1}+\frac{dy^{2}}{1-y^{2}}\right)\,,
\end{eqnarray}
where the metric functions $U$ and $V$ depend only on the prolate spheroidal coordinates $x \in [1, +\infty)$ and $y\in [-1,1]$. The coordinate $t$ parameterizes time translations, and the coordinate $\phi$ describes rotations around the axis of symmetry. The solution contains a Killing  horizon located at $x=1$, while the symmetry axis corresponds to $y = \pm 1$. The real parameter $m$ is equal to the Komar mass on the horizon. The prolate spheroidal coordinates are related to the  Schwarzschild coordinates $r$ and $\theta$ as

\begin{eqnarray}
x=\frac{r}{m}-1, \quad~~~ y=\cos\theta\,. \nonumber
\end{eqnarray}

The metric functions  $U$ and $V$  possess the form
\begin{eqnarray}
U&=&\sum_{n\geq0}a_{n}R^{n}P_{n}\,,\\
V&=&\sum_{n,k\geq1}\frac{nka_{n}a_{k}}{(n+k)}R^{n+k}(P_{n}P_{k}-P_{n-1}P_{k-1})\,\nonumber\\
&+&\sum_{n\geq1}a_{n}\sum_{l=0}^{n-1}[(-1)^{n-l+1}(x+y)-x+y]R^{l}P_{l}\,, \nonumber\\
P_{n}&\equiv&P_{n}(xy/R), \quad~~~ R=\sqrt{x^{2}+y^{2}-1},  \nonumber
\end{eqnarray}
where $P_n$ are the Legendre polynomials, and $a_n$ are real constants. Due to their presence the solution is not asymptotically flat if considered as global solution. If we consider a local solution, they are interpreted as encoding the influence of a gravitational source located in the exterior of the region of validity of ($\ref{metric}$). In the limit when all the constants $a_n$ vanish, we obtain the asymptotically flat Schwarzschild solution. The function $U(x,y)$ is a harmonic function defined in a nonphysical 3D flat space. In analogy with the terminology used in the Newtonian gravity and electromagnetism, the coefficients $a_n$ are called multipole moments, since they should coincide with the interior multipole moments in the multipole expansion of the external gravitational field. Constraining the values of the multipole moments, we impose restrictions on the type and symmetries of the external gravitational source.

The distorted Schwarzschild solution is free of conical singularities on the symmetry axis $y = \pm 1$ provided the multipole moments satisfy the condition
\begin{eqnarray}
\sum_{n\geq0}a_{2n+1}=0\,.
\end{eqnarray}
In this case the metric can describe balanced configurations of a black hole and external matter.

For our purposes it is convenient to present the metric in the form
\begin{eqnarray}
dS^{2}&=&-\left(\frac{x-1}{x+1}\right)e^{2{\cal U}}dt^{2}+(x+1)^{2}(1-y^{2})e^{-2{\cal U}}d\phi^{2}\nonumber\\
&+&(x+1)^{2}e^{2(V-{\cal U})}\left(\frac{dx^{2}}{x^{2}-1}+\frac{dy^{2}}{1-y^{2}}\right)\,,
\end{eqnarray}
where we introduce the notations
\begin{eqnarray}
{\cal U}=U-u_{0}, \quad~~~ u_{0}=\sum_{n\geq0}a_{n}\, . \nonumber
\end{eqnarray}

Thus, the metric is rescaled by a conformal factor, and the time coordinate is redefined as
\begin{eqnarray}
ds^{2}=\Omega^{2}dS^{2}, \quad~~~\Omega^{2}=m^{2}e^{-2u_{0}}\,,\quad~~~t\rightarrow \frac{1}{m}e^{2u_{0}}t. \nonumber
\end{eqnarray}
As a result, the new metric is characterized by a unit Komar mass on the horizon, and the horizon area is equal to $16\pi$ as for the asymptotically flat Schwarzschild solution with unit mass.

In the following we will consider a special type of distortion, in which only the quadrupole moment is nonzero, while all the others vanish, i.e. $\{a_2 = q\neq 0; a_n = 0,  n\neq 2\}$. Then, the metric functions obtain the form

\begin{eqnarray}
{\cal U} &=& \frac{1}{2}q(3x^2y^2 - x^2 - y^2 -1)\, ,  \\
V &=& 2qx(y^2-1) \nonumber \\
&+& \frac{1}{4}q^2\left[9x^4y^4 + (x^2+y^2-1)(x^2+y^2-1-10x^2y^2)\right]\, . \nonumber
\end{eqnarray}

\section{Circular orbits in the equatorial plane}

\begin{figure}[t!]
        \centering			
           \includegraphics[width=0.5\textwidth]{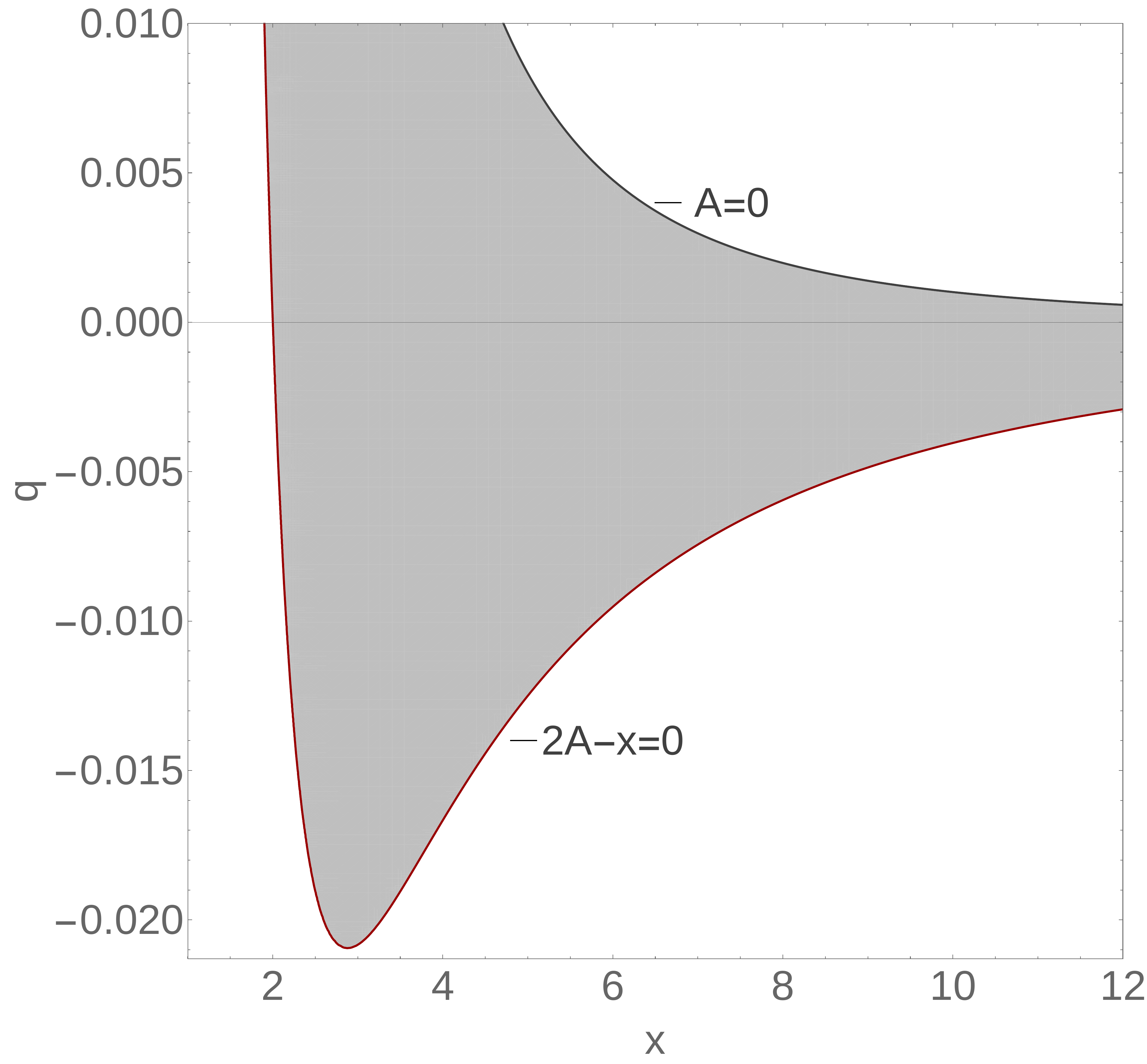}
		    \caption{\label{fig:existence}\small Domain of existence of the circular orbits in the equatorial plane. Null orbits are located on the curve $2A-x=0$ represented in red. Timelike orbits belong to the region bounded by the curves $2A-x=0$ and $A=0$ depicted in gray.}
\end{figure}

We consider the geodesic equations for the distorted Schwarzschild solution
\begin{equation}\label{geodesics}
\ddot{x^{\alpha}}+\Gamma^{\alpha}_{\,\,\,\beta\gamma}\dot{x^{\beta}}\dot{x^{\gamma}}=0\,,
\end{equation}
where $x^{\alpha}= \{t, x, y, \phi\}$ are the spacetime coordinates, the dot denotes differentiation with respect to an affine parameter, and $\Gamma^{\alpha}_{\,\,\,\beta\gamma}$ are the Christoffel symbols. We give the explicit form of the connection coefficients for our solution in the Appendix. Due to the spacetime symmetries the geodesic motion is characterized by two conserved quantities ${\cal E}$ and ${\cal L}$

\begin{eqnarray}
{\cal E}&=&\left(\frac{x-1}{x+1}\right)e^{2{\cal U}}\,\dot{t}\,, \nonumber\\
{\cal L} &=& (x+1)^{2}(1-y^{2})e^{-2{\cal U}}\dot{\phi}\,,
\end{eqnarray}
which are associated with the particle's specific energy and angular momentum. We further have the constraint equation $g_{\mu\nu}{\dot x}^{\mu}{\dot x}^{\nu} =\epsilon$, where $\epsilon$ takes the value $\epsilon = -1$ for timelike geodesics, and $\epsilon = 0$ for null geodesics. If we restrict the motion to the equatorial plane $y=0$, the constraint equation takes the form

\begin{eqnarray}
e^{2V}\dot{x}^{2}&=&{\cal E}^{2}-{\cal U}_{eff}\,,\nonumber \\
{\cal U}_{eff}&=&\left(\frac{x-1}{x+1}\right)e^{2{\cal U}}\left[-\varepsilon+\frac{{\cal L}^{2}\,e^{2{\cal U}}}{(x+1)^{2}}\right]\,,
\end{eqnarray}
where we introduce the effective potential ${\cal U}_{eff}$. The properties of the effective potential and the qualitative behavior of the equatorial geodesic motion were studied in detail in \cite{Shoom:2016}. The possible positions of the circular orbits can be obtained by determining the stationary points of the effective potential. For null geodesics they are located on the curve

\begin{eqnarray}\label{exist_null}
2-x -2qx(x^{2}-1) = 0.
\end{eqnarray}
The position of the circular orbits does not depend on the photon's angular momentum, and in the limit $q=0$ it reduces to the location of the photon sphere for the Schwarzschild solution $x=2$. For positive quadrupole moments there is a single circular orbit for every $q$ (see fig. $\ref{fig:existence}$). For negative quadrupole moments the curve ($\ref{exist_null}$) is bounded and reaches a minimum at $q_{min}\approx -0.021$, $x_{min}\approx 2.879$. Therefore, there exists a minimum quadrupole moment, for which circular orbits can exist. For $q>q_{min}$ there are two circular orbits, which deviate from each other when the quadrupole moment increases. In the limit $q\rightarrow0$ one of them approaches the location of the photon sphere for the Schwarzschild solution $x=2$, while the other one tends to infinity.

For timelike geodesics the positions of the circular orbits are determined by the solutions to the following algebraic equation \cite{Shoom:2016}

\begin{eqnarray}\label{extr_timelike}
\frac{(x+1)^{2}[1-qx(x^{2}-1)]}{[x-2+2qx(x^{2}-1)]}={\cal L}^{2}e^{-q(x^{2}+1)}\,.
\end{eqnarray}

They form a region in the $(x, q)$-plane bounded by the locus of the circular orbits for the null geodesics

\begin{eqnarray}
2-x -2qx(x^{2}-1) = 0\,, \nonumber
\end{eqnarray}
and the curve
\begin{eqnarray}
1 - qx(x^{2}-1) =0\,,
\end{eqnarray}
which is illustrated in fig. $\ref{fig:existence}$. It is convenient to introduce the notation
\begin{eqnarray}
A = 1 - qx(x^{2}-1).
\end{eqnarray}
Then the  domain of existence of the circular timelike geodesics is determined by the inequalities $A>0\,\cap \, 2A-x <0$. For every quadrupole moment $q>q_{min}$ the radial positions of the circular orbits belong to a certain interval $(x_{-}, x_{+})$. The lower limit $x_{-}$ corresponds to the position of the circular photon orbit for the particular quadrupole moment, which lies closer to the horizon.  When the absolute value of $q$ decreases, the upper limit $x_{+}$ increases, and in the limit $q\rightarrow0$ it tends to infinity. Thus, the Schwarzschild limit is reproduced with circular timelike orbits existing in the interval $x\in[2, +\infty)$.

While for $q<0$ circular orbits are possible only above the quadrupole moment $q_{min}\approx -0.021$, they can exist for any positive value of the quadrupole moment. When $q>0$ increases, the interval $(x_{-}, x_{+})$ decreases, and its lower limit approaches the horizon $x=1$. In the limit $q\rightarrow +\infty$ both curves $A=0$ and $2A-x =0$ tend to the location of the horizon.

Each circular orbit is characterized by the orbital frequency $\omega_0$ of the particles moving on it, and their specific energy ${\cal E}$ and angular momentum ${\cal L}$. They can be calculated using the general formula for any static and axisymmetric metric

\begin{eqnarray}
{\cal E}&=&-\frac{g_{tt}}{\sqrt{-g_{tt}-g_{\phi\phi}\omega_0^2}},    \label{rotE}  \\
{\cal L}&=&\frac{g_{\phi\phi}\omega_0}{\sqrt{-g_{tt}-g_{\phi\phi}\omega_0^2}},     \label{rotL}  \\
\omega^2_0&=&\left(\frac{d\phi}{dt}\right)^2=-\frac{g_{tt,x}}{g_{\phi\phi,x}}.     \label{rotOmega}
\end{eqnarray}

In the case of the distorted Schwarzschild solution the kinematic quantities take the explicit form

\begin{eqnarray}\label{omega_0}
\omega^2_0 &=& \frac{e^{4\cal U}}{(x+1)^3}\frac{1-qx(x^2-1)}{1+qx(x+1)} \nonumber \\
 &=& \frac{e^{4\cal U}}{(x+1)^3}\frac{A(x-1)}{x-A}, \\ \nonumber
{\cal E} &=& (x+1) e^{-\cal U}\sqrt{\frac{A}{(x-2A)}}\,, \\ \nonumber
{\cal L} &=& \pm (x+1)e^{-\cal U}\sqrt{\frac{A(x-1)}{(A-x)(2A-x)}}.
\end{eqnarray}

\section{Stability of the circular orbits}

In order to investigate the stability of the circular orbits in the equatorial plane in the linear approximation we consider a small deviation from the circular motion $\tilde{x}^{\mu}(s) = x^{\mu}(s) + \xi^\mu(s)$ , where $x^{\mu}(s)$ describes the circular orbit, and $s$ is an affine parameter on the particle trajectory. We substitute this expression in the geodesic equations $(\ref{geodesics})$ and considering terms up to linear order in $\xi^\mu(s)$, we obtain the following system \cite{Aliev:1981}, \cite{Aliev:1986}

\begin{eqnarray}\label{pert}
&&\frac{d^2\xi^\mu}{dt^2} + 2\gamma^\mu_\alpha\frac{d\xi^\alpha}{dt} + \xi^b\partial_b{\cal V}^{\mu} = 0\, , \quad b = x,y \nonumber \\[2mm]
&&\gamma^\mu_\alpha =\left[\Gamma^\mu_{\alpha\beta} u^\beta(u^0)^{-1}\right]_{y=0}\, , \nonumber \\[2mm]
&& {\cal V}^{\mu} = \left[\gamma^\mu_\alpha u^\alpha(u^0)^{-1}\right]_{y=0},
\end{eqnarray}
which describes the dynamics of the small perturbation. In these equations we express the 4-velocity vector for the circular orbits in the equatorial plane as $u^\mu = \dot{x^\mu}= u^0(1, 0, 0, \omega_0)$, where $u^0 = dt/ds$ and $\omega_0$ is the orbital frequency $(\ref{omega_0})$. We further introduce the convention that Greek indices run over all the spacetime coordinates, capital Latin indices refer to the cyclic coordinates $t$ and $\phi$, while small Latin indices denote the prolate coordinates $x$ and $y$.  The equations for the $t$ and $\phi$ components in $(\ref{pert})$ can be integrated leading to

\begin{eqnarray}
\frac{d\xi^A}{dt} + 2\gamma^A_\alpha\xi^\alpha = 0\, , \quad A = t,\phi.
\end{eqnarray}

Substituting these relations in the remaining part of the system, we can show that for any static and axisymmetric solution the equations for the radial and vertical perturbations decouple and reduce to the form
\begin{eqnarray}
&&\frac{d^2\xi^x}{dt^2} + \omega_x^2\xi^x = 0\, , \label{pertx} \\[2mm]
&&\frac{d^2\xi^y}{dt^2} + \omega_y^2\xi^y = 0\, ,  \label{perty}
\end{eqnarray}
where we introduce the quantities
\begin{eqnarray}
&&\omega_x^2 = \partial_x {\cal V}^x - 4\gamma^x_A\gamma^A_x, \\[2mm] \nonumber
&&\omega_y^2 = \partial_y{\cal V}^y. \nonumber
\end{eqnarray}
The dynamics of the perturbations in the radial and the vertical direction is determined by the sign of  $\omega_x^2$ and $\omega_y^2$. For positive values eqs. $(\ref{pertx})$-$(\ref{perty})$ describe a couple of harmonic oscillators, i.e.  small deviations from the circular orbits will oscillate in their vicinity with frequencies $\omega_x$ and $\omega_y$ in the radial and vertical direction,  respectively. In this case the circular orbit is stable in the linear approximation, and the quantities $\omega_x$ and $\omega_y$ are called epicyclic frequencies. If one of the quantities $\omega_x^2$ and $\omega_y^2$ is negative, the circular orbit is unstable, since small perturbations in the corresponding direction will deviate exponentially from it.

Performing the calculations for the distorted Schwarzschild solution,  we obtain the following expressions for the epicyclic frequencies

\begin{eqnarray}
\omega^2_y&=& \frac{\omega_0^2e^{-2V}}{A}(1+2qx^3)\, ,  \\[2mm]
\omega^2_x&=& \frac{\omega_0^2e^{-2V}}{x^2-1}\left[2(2A-x)(A-x) -\frac{(x^2-1)}{A}(1+2qx^3)\right]\, , \nonumber \\
A &=& 1-qx(x^2-1)\, , \nonumber
\end{eqnarray}
by means of the orbital frequency $\omega_0$ given by eq. $(\ref{omega_0})$. We should note that the epicyclic frequencies, as well as all the kinematic properties of the circular orbits, are well-defined only in the region of existence of the circular orbits defined by the conditions $A>0\cap 2A-x <0$. We can further write  the relation
\begin{eqnarray}
\omega^2_x + \omega^2_y &=& \frac{2\omega_0^2e^{-2V}}{x^2-1}(2A-x)(A-x)\, ,
\end{eqnarray}
which shows that in the region of existence of the circular orbits the sum $\omega^2_x + \omega^2_y$ is always non-negative.

The stability condition for the circular orbits with respect to perturbations in the radial direction also coincides with the requirement that they are located at the minima of the effective potential ${\cal U}_{eff}$. The derivative $\partial^2_{xx}{\cal U}_{eff}$ evaluated at the stationary points of the effective potential is given by the expression

\begin{eqnarray}\label{Vxx}
\partial^2_{xx}{\cal U}_{eff}&=& \frac{2e^{2{\cal U}}}{(x^2-1)(x+1)^2(x-2A)}\bigg[2(2A-x)(A-x) \nonumber\\[2mm]
&-&\frac{(x^2-1)}{A}(1+2qx^3)\bigg]\, ,
\end{eqnarray}
which is positive for positive values of  $\omega^2_x$.

\begin{figure}[t!]
    		\setlength{\tabcolsep}{ 0 pt }{\small\tt
		\begin{tabular}{ cc}
           \includegraphics[width=0.5\textwidth]{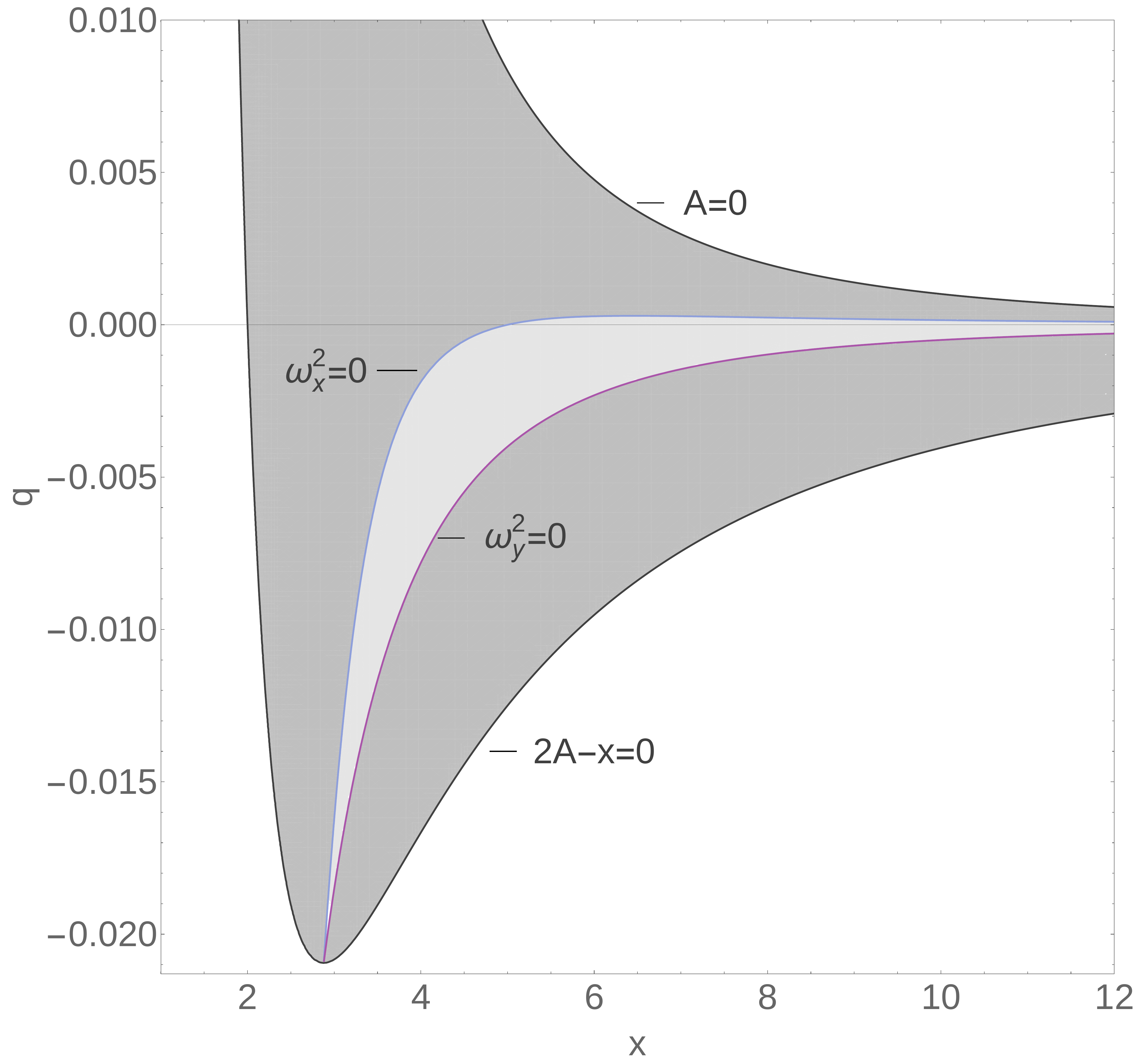}
		   \includegraphics[width=0.5\textwidth]{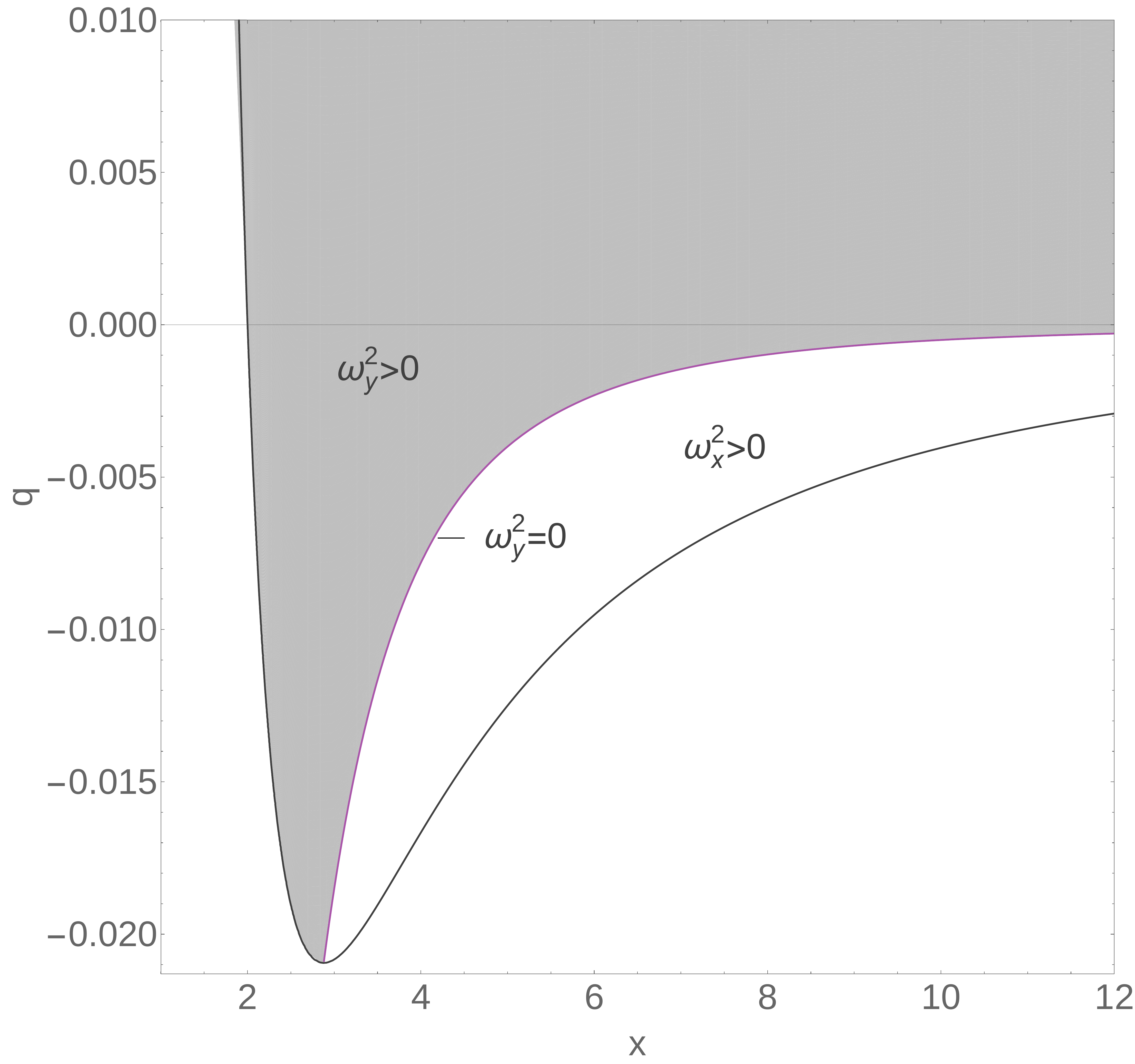}
                   \end{tabular}}
 \caption{\label{fig:stability}\small Stability of the timelike (left) and null (right) circular orbits for the distorted Schwarzschild black hole. Timelike circular orbits are stable in the light gray region bounded between the blue and the purple curves, which represent the relations $\omega^2_x = 0$ and $\omega^2_y = 0$, respectively.  For null circular orbits we plot only the curve $\omega^2_y = 0$, since we have $\omega^2_x =-\omega^2_y$. In the region above the purple curve the orbits are stable with respect to vertical perturbations and unstable with respect to radial perturbations, while in the region below the curve the opposite is satisfied.}
\end{figure}

In fig. $\ref{fig:stability}$ we present the analysis of the stability of the timelike circular orbits with respect to linear perturbations. The domain of existence is illustrated by the dark gray area bounded between the curves  $A=0$ and $2A-x=0$. We further plot the curves $\omega^2_x = 0$ and $\omega^2_y = 0$, which are depicted in blue and purple, respectively. The circular orbits are stable with respect to radial perturbations in the region below the blue curve, where $\omega^2_x > 0$ is satisfied. On the contrary,  stability with respect to vertical perturbations is achieved above the purple curve, where we have $\omega^2_y > 0$. Thus, both the stability conditions are fulfilled in the region bounded between the two curves where both the inequalities are positive. Consequently, for values of the parameters $q$ and $x$ belonging to this region, the circular orbits are stable with respect to both radial and vertical perturbations, and $\omega_x$ and $\omega_y$ can be interpreted as epicyclic frequencies of oscillation around the circular equatorial motion.

The curves $\omega^2_x = 0$ and $\omega^2_y = 0$ intersect at a single point $P = \{ x_{min}\approx 2.879, q_{min}\approx -0.021\}$, which corresponds to the minimal quadrupole moment for which circular photon orbits can exist. Thus, this point represents also the lower limit for the location of the stable timelike orbits. The curve $\omega^2_x = 0$ crosses the axis $q=0$ at the location of the ISCO for the Schwarzschild solution $x=5$, while $\omega^2_y = 0$ has no intersection points. Instead, it approaches the axis $q=0$ asymptotically in the limit $x\rightarrow\infty$. Consequently, for positive quadrupole moments $\omega^2_y$ is always positive, and the stability of the orbit is determined only by its stability with respect to radial perturbations. The curve $\omega^2_x = 0$ has a maximum located at $Q = \{ x_{max}\approx 6.47 , q_{max}\approx 0.00027\}$, which determines the maximal value of the quadrupole moment for which stable circular orbits can exist. For large $x$ it tends asymptotically to the axis $q=0$.

A significant difference from the isolated Schwarzschild black hole is that the region of stability is bounded in the radial direction. For the Schwarzschild solution the vertical epicyclic frequency is always positive and the stability is determined only by the radial epicyclic frequency. Thus, all the orbits with radial positions larger than the ISCO are stable. For the distorted Schwarzschild black hole with positive quadrupole moment again only the radial epicyclic frequency determines the stability. However, there are two marginally stable orbits with respect to perturbations in the radial direction. For negative quadrupole moment the requirement that the orbits are stable with respect to vertical perturbations imposes an upper limit on the region of stability. Consequently, the stable orbits both for positive and negative distortion are located in an annular region bounded between two marginally stable orbits.

Since the distorted Schwarzschild black hole is a local solution, it is supposed to be valid up to a certain radial distance after which it should be extended in order to construct a global solution. Depending on the physical scenario, which the global solution should describe, the stable circular orbits can be distributed in a different way. This imposes restrictions on the location of the boundary of the region of validity. If we require that the global solution possesses a continuous region of stable circular orbits spanning from a certain ISCO to infinity, the boundary of the region of validity of the distorted Schwarzschild solution should be chosen within the region of stability of the circular orbits. Then, by extending the distorted black hole to an appropriately chosen exterior solution,  it is possible to achieve an unbounded distribution of stable circular orbits. In this situation the outer marginally stable orbit for a fixed value of the quadrupole moment sets an upper limit for the possible region of validity of the distorted solution.

However, certain compact objects naturally possess annular regions of stability of the circular orbits. Examples for this property are some naked singularities like the Janis-Newman-Winicour solution or the Reissner-Nordstr\"{o}m solution. If we aim at a global solution with such a distribution of the stable circular orbits, the boundary of the region of validity for the distorted Schwarzschild solution can be chosen also outside the region of stability.

\subsection{Null circular orbits}

We can obtain the epicyclic frequencies for the null circular orbits as a limit of the timelike case. The existence condition for circular null orbits is given by the equation $2A-x =0$. Imposing this condition on the epicyclic frequencies for the timelike circular orbits, we  obtain the corresponding expressions  for the null geodesics

\begin{eqnarray}\label{omega_null}
\omega^2_y= -\omega^2_x = \frac{\omega_0^2e^{-2V}}{A}(1+2qx^3)\, .
\end{eqnarray}
We see that the radial and the vertical epicyclic frequencies are not independent, and always possess opposite signs. Therefore, orbits which are stable with respect to radial perturbations will be unstable with respect to vertical ones, and vice versa.  The region of stability with respect to radial and vertical perturbations is presented in fig. \ref{fig:stability}. The circular orbits are located on the curve $2A-x=0$, and the curve $\omega^2_y = 0$ is depicted in purple. The grey area corresponds to positive values of $\omega^2_y$, and consequently negative values of $\omega^2_x$. For positive quadrupole moments all the circular orbits are unstable  with respect to radial perturbations, and stable for vertical  ones. For negative quadrupole moments the circular orbits, which are located closer to the black hole horizon are unstable with respect to radial perturbations, while the more distant one is stable. They correspond, respectively, to maxima and minima of the effective potential ${\cal U}_{eff}$, which determines the motion in the equatorial plane. When the quadrupole moment decreases, the two orbits approach each other and finally merge at the point $P = \{ x_{min}\approx 2.879, q_{min}\approx -0.021\}$, in which the region of existence $2A-x=0$ intersects the curve $\omega^2_y = 0$. If we look at the alternative description by means of the effective potential, this point corresponds to its inflection point.

The stability of the circular null geodesics was studied previously also in \cite{Nedkova:2018}-\cite{Shoom:2017}. It was demonstrated that by taking advantage of eq. ($\ref{exist_null}$) the orbital and epicyclic frequencies can be expressed as functions only of the position of the circular orbit in the form

\begin{eqnarray}
\omega^2_0 &=& \frac{x-1}{(x+1)^3}\exp\left[\frac{x(x-2)}{x^2-1}\right]\, ,  \\[2mm]
\omega^2_x &=& -\omega^2_y = \frac{2\omega_0^2\,e^{-2V}}{x(x^2-1)}(x^3 -3x^2 +1)\, , \nonumber \\[2mm]
V&=&\frac{2-x}{16x^2(x^2-1)}(x^3-18x^2-x+2)\,. \nonumber
\end{eqnarray}

\subsection{Properties of the epicyclic frequencies}

\begin{figure}[t!]
    		\setlength{\tabcolsep}{ 0 pt }{\small\tt
		\begin{tabular}{ cc}
           \includegraphics[width=0.5\textwidth]{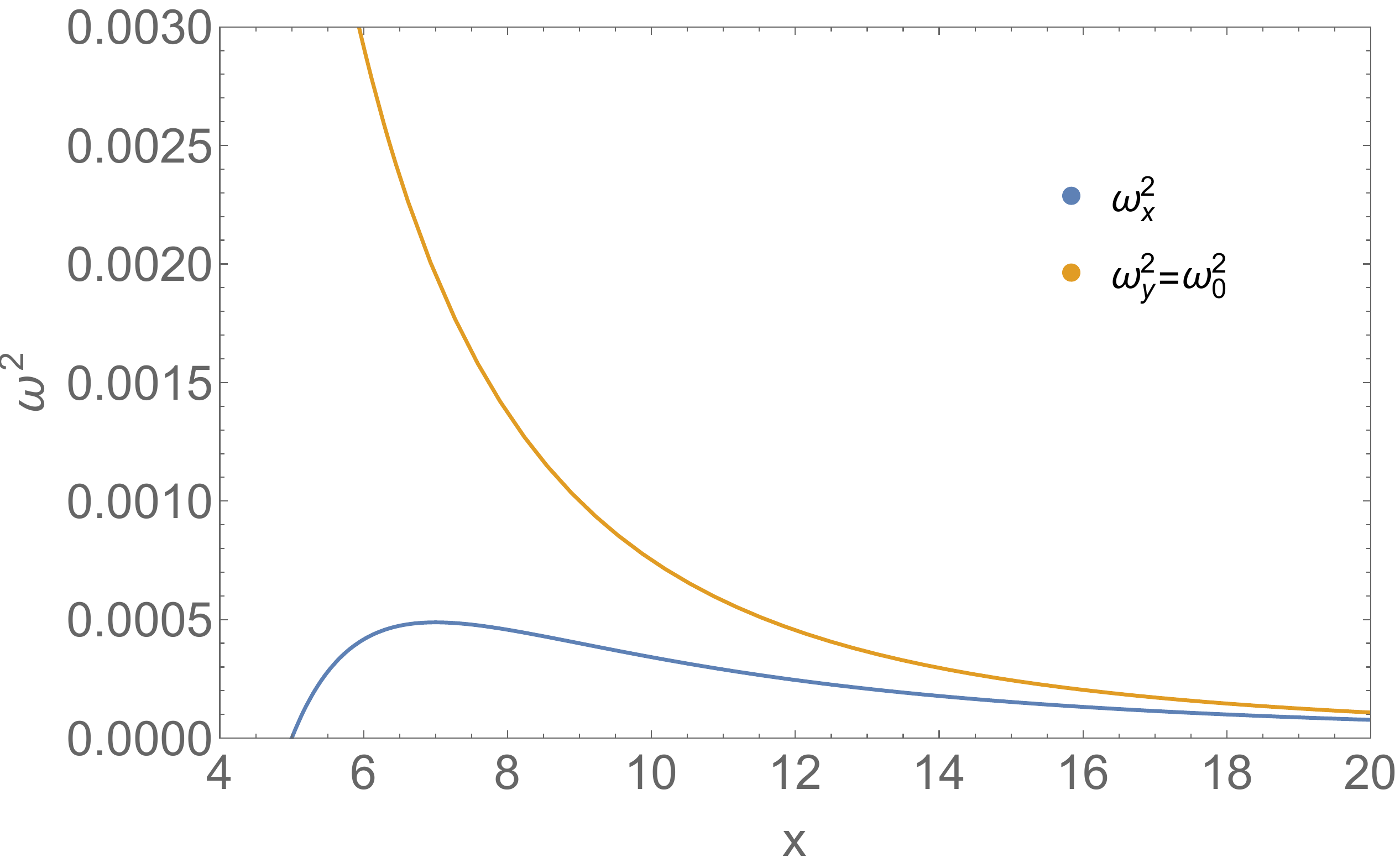}
		   \includegraphics[width=0.5\textwidth]{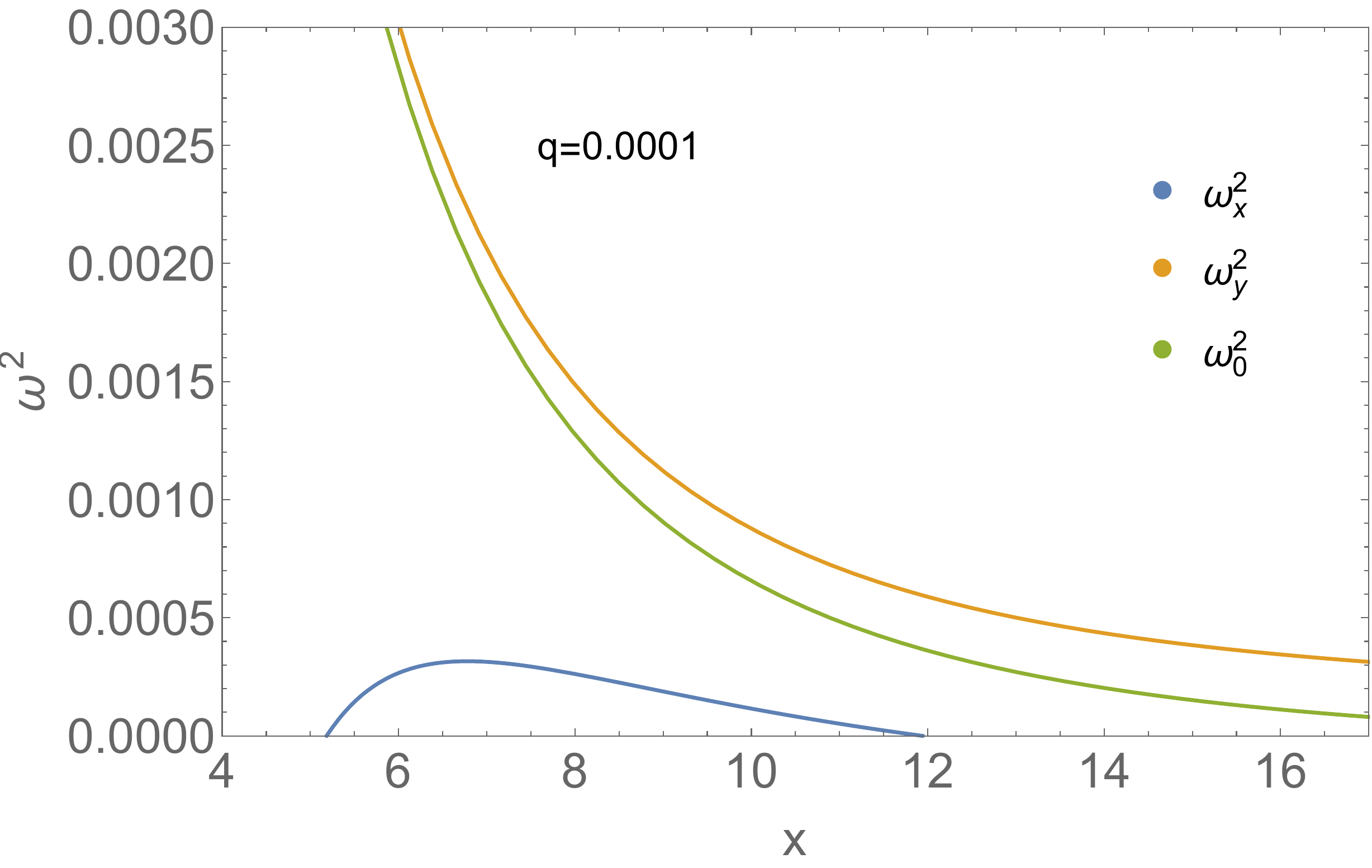} \\[1mm]
           \hspace{1.2cm}  $a)$ \hspace{7cm}  $b)$ \\[3mm]
           \includegraphics[width=0.5\textwidth]{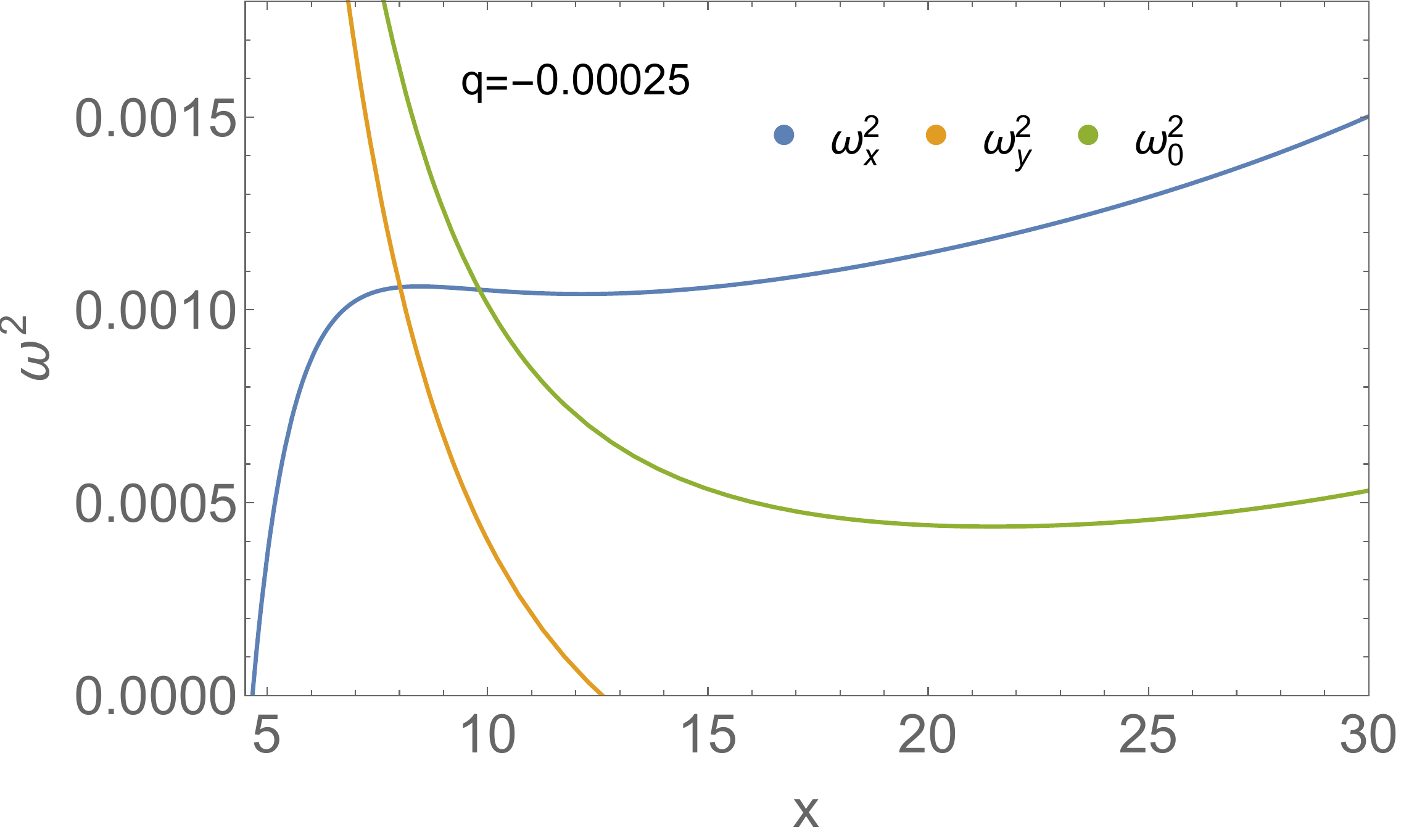}
           \includegraphics[width=0.5\textwidth]{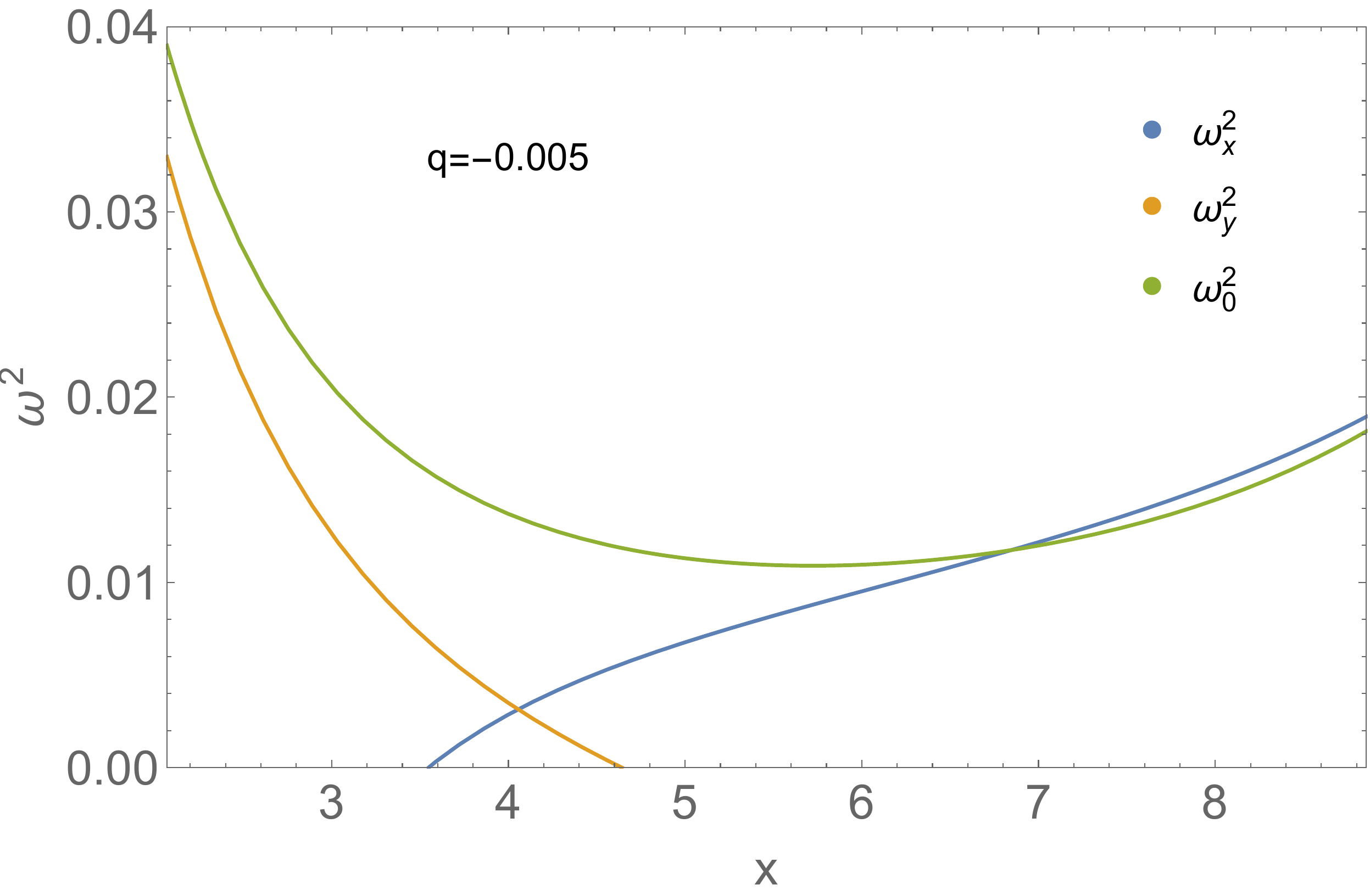}\\[1mm]
           \hspace{1.2cm}  $c)$ \hspace{7cm}  $d)$
        \end{tabular}}
 \caption{\label{fig:frequency}\small The possible qualitatively different types of behaviour of the epicyclic and the orbital frequencies for the distorted Schwarzschild solution. The case of the Schwarzschild black hole is also presented for comparison in  a). }
\end{figure}

In this section we examine some of the properties of the epicyclic frequencies for the timelike circular orbits, which are relevant for the occurrence of  different kinds of non-linear resonances, and we compare them to the case of the Schwarzschild solution. For the Schwarzschild black hole the epicyclic frequencies are given by the expressions

\begin{eqnarray}
\omega^2_y &=& \omega^2_0 = \frac{M}{r^3}, \nonumber\\
\omega^2_x &=& \omega^2_0\left(1-\frac{6M}{r}\right),
\end{eqnarray}
in the usual Schwarzschild coordinates. The stability of the circular orbits is determined only by the radial epicyclic frequency, since the vertical one coincides with the orbital frequency $\omega_0$ and is always positive. Therefore, the stable circular orbits are located at radial distances larger that the location of the ISCO at $r_{ISCO}=6m$, or in the prolate spheroidal coordinates $x_{ISCO}=5$, and the unstable ones belong to the interval $(x_{ph}, x_{ISCO})$, where $x_{ph} =2$ is the position of the photon sphere. The radial epicyclic frequency has a single maximum. The vertical epicyclic frequency is a monotonically decreasing function and it is always larger than the radial one (see fig. $\ref{fig:frequency}$ a)). This ordering between the frequencies imposes restrictions on the possible resonances that could occur.

\begin{figure}[t!]
    		\setlength{\tabcolsep}{ 0 pt }{\small\tt
		\begin{tabular}{ cc}
           \includegraphics[width=0.5\textwidth]{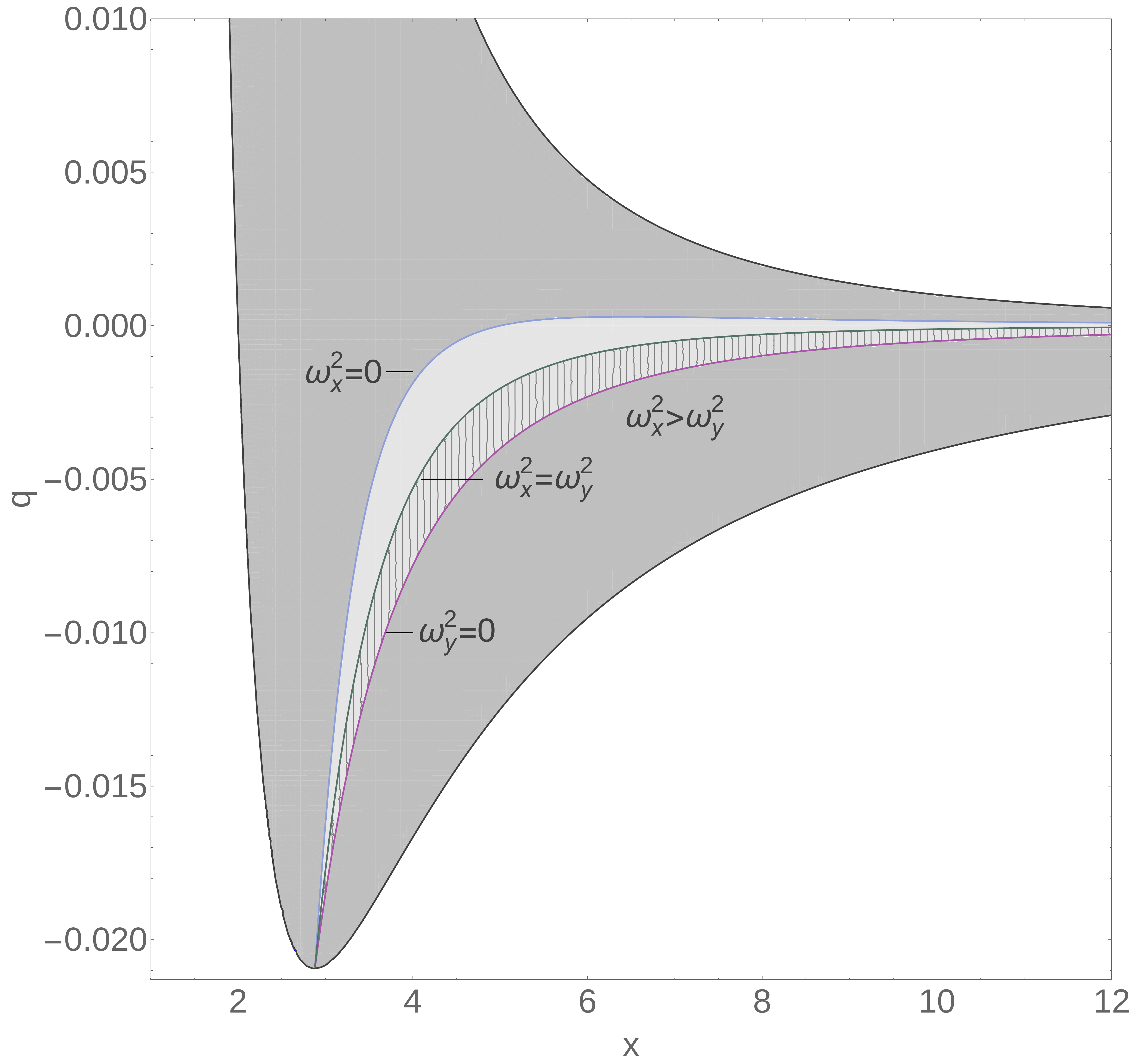}
		   \includegraphics[width=0.5\textwidth]{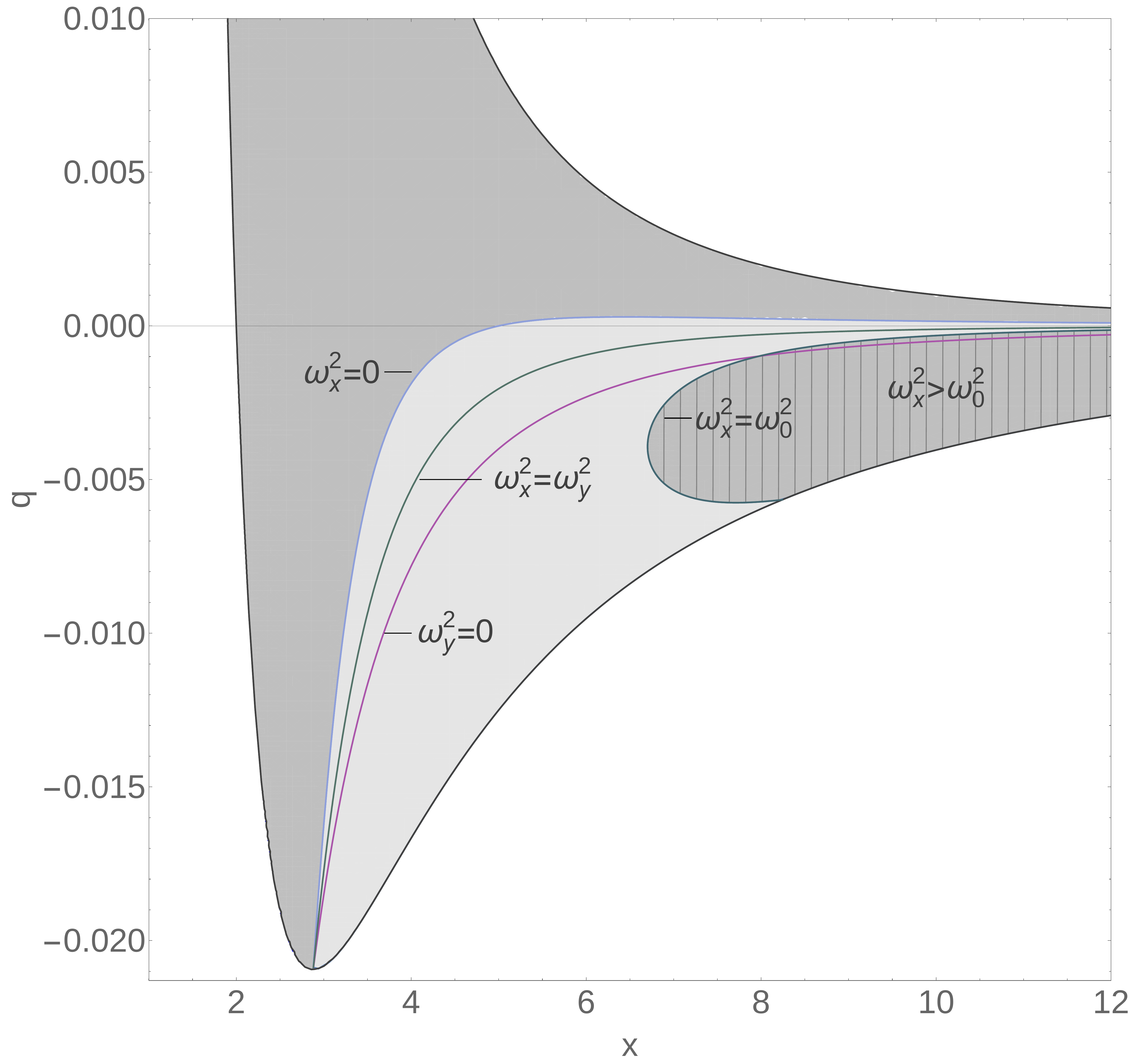} \\[2mm]
            \includegraphics[width=0.5\textwidth]{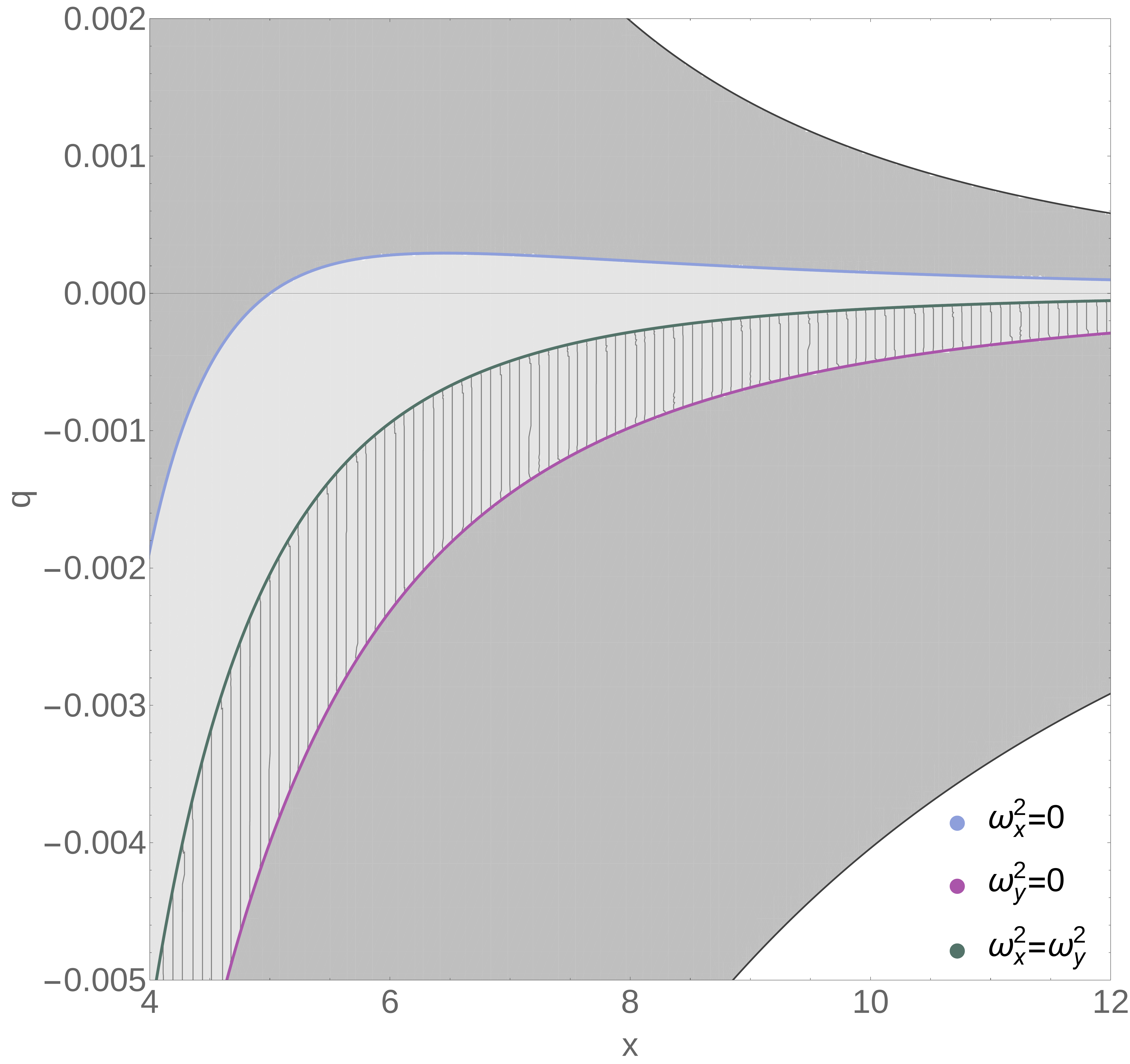}
		     \includegraphics[width=0.5\textwidth]{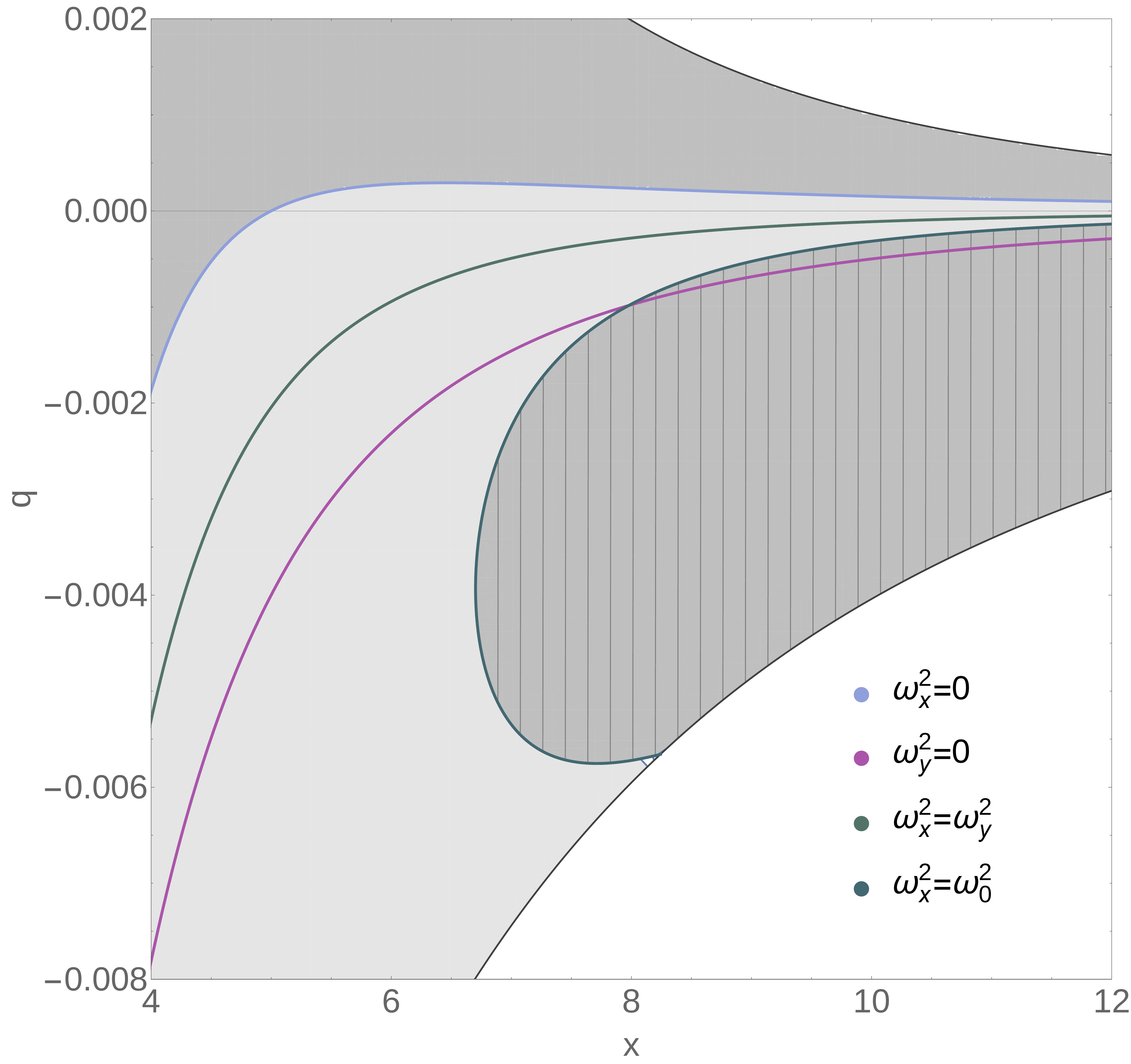}
                   \end{tabular}}
 \caption{\label{fig:omega_xy}\small Comparison between the epicyclic and the orbital frequencies. Left column: The curve $\omega^2_x = \omega^2_y$ is plotted in green. The  curves $\omega^2_x = 0$ and  $\omega^2_y= 0$ are plotted in blue and purple respectively, and both the epicyclic frequencies are positive in the light gray region between them. In the region below the green curve it is satisfied that $\omega^2_x > \omega^2_y$, and the hatched area denotes its intersection with the region of stability of the circular motion. The domain of existence of the circular orbits is represented in dark gray. Right column: The curve $\omega^2_x = \omega^2_0$ is plotted in dark green. In the hatched region  below the curve it is satisfied that $\omega^2_x > \omega^2_0$. }
\end{figure}

In the case of the distorted Schwarzschild black hole much more diverse situations can be realized. All the characteristic frequencies are independent, and there exist regions in the parametric space with different kinds of ordering between them. The various cases that can occur are investigated in fig. $\ref{fig:omega_xy}$ , where we compare the three frequencies by plotting the curves $\omega^2_x = \omega^2_y$, and $\omega^2_x = \omega^2_0$. Below the curve $\omega^2_x = \omega^2_y$ we have $\omega^2_x > \omega^2_y$, and in a similar way below the curve $\omega^2_x = \omega^2_0$ it is satisfied that $\omega^2_x > \omega^2_0$.  The solutions of the equation $\omega^2_y = \omega^2_0$ coincide with the axis $q=0$, as $\omega^2_y >\omega^2_0$ is fulfilled for $q>0$. The curves $\omega^2_x = \omega^2_y$, and $\omega^2_x = \omega^2_0$ approach asymptotically the axis $q=0$ for large $x$ without crossing it. Hence, for positive quadrupole moments the three characteristic frequencies are ordered as $\omega^2_y > \omega^2_0>\omega^2_x$. For negative quadrupole moments three different cases are possible. In the region above the curve $\omega^2_x = \omega^2_y$ up to the axis $q=0$ we have the ordering $\omega^2_0 > \omega^2_y>\omega^2_x$. Between the curves $\omega^2_x = \omega^2_y$ and $\omega^2_x = \omega^2_0$ the frequencies satisfy $\omega^2_0 > \omega^2_x>\omega^2_y$, while below the curve $\omega^2_x = \omega^2_0$ the inequality $\omega^2_x > \omega^2_0>\omega^2_y$ is valid. We note that all the analysis of the epicyclic frequencies should be constrained to the domain of existence of the timelike circular orbits, which is denoted by the dark gray area in the figures.

We further investigate the behavior of the epicyclic and the orbital frequencies as a function of the radial distance $x$. Keeping the quadrupole moment as a parameter we
evaluate their possible form for a fixed value of $q$. Unlike the Schwarzschild solution where $\omega_y=\omega_0$ is monotonically decreasing, while $\omega_x$ has a single maximum, we can observe qualitatively different cases depending on the range of the quadrupole moment. The vertical epicyclic frequency $\omega_y$ is always monotonically decreasing  for any value of the quadrupole moment in the domain of existence of the timelike circular orbits. The radial epicyclic frequency $\omega_x$ can have a single maximum, a maximum and a minimum, or no extrema in the various ranges of $q$. The orbital frequency $\omega_0$ either has no extrema, or possesses a single minimum. Thus, we observe the following three situations in the domain of existence of the timelike circular orbits.

\begin{itemize}
\item $q\geq 0$: $\omega_y$ and $\omega_0$ are monotonically decreasing, and $\omega_x$ has a single maximum.
\item $0>q>q_{crit} \approx -0.00034$:  $\omega_y$ is monotonically decreasing, $\omega_0$ has a single minimum, and $\omega_x$ possesses a minimum and a maximum.
\item $q<q_{crit}$: $\omega_y$ is monotonically decreasing, $\omega_x$ is monotonically increasing, while $\omega_0$ has a single minimum.
\end{itemize}

\begin{figure}[t!]
    		\setlength{\tabcolsep}{ 0 pt }{\small\tt
		\begin{tabular}{ cc}
           \includegraphics[width=0.5\textwidth]{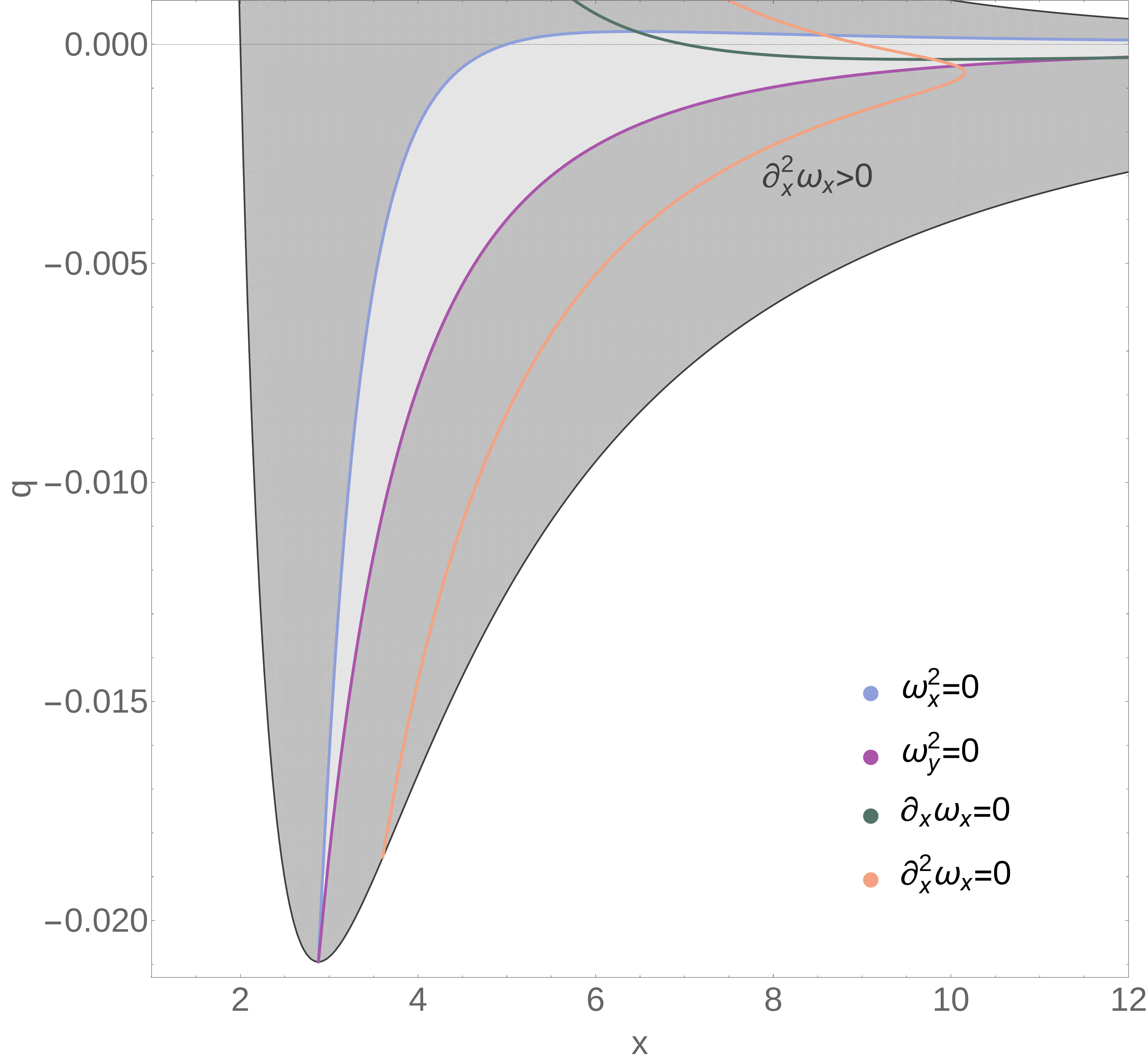}
		   \includegraphics[width=0.5\textwidth]{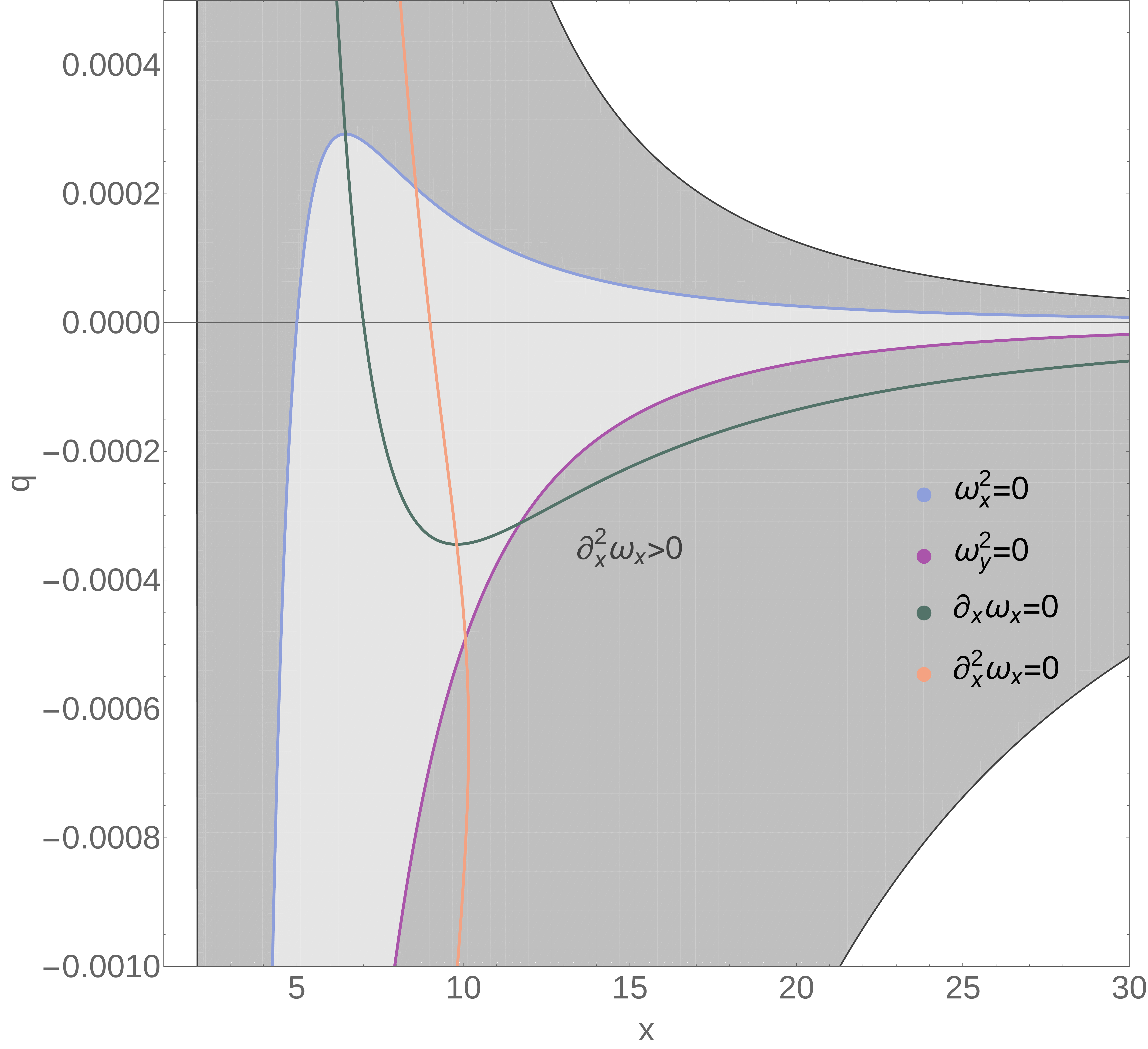} \\[1mm]
           \hspace{1.2cm}  $a)$ \hspace{7cm}  $b)$ \\[3mm]
           \includegraphics[width=0.5\textwidth]{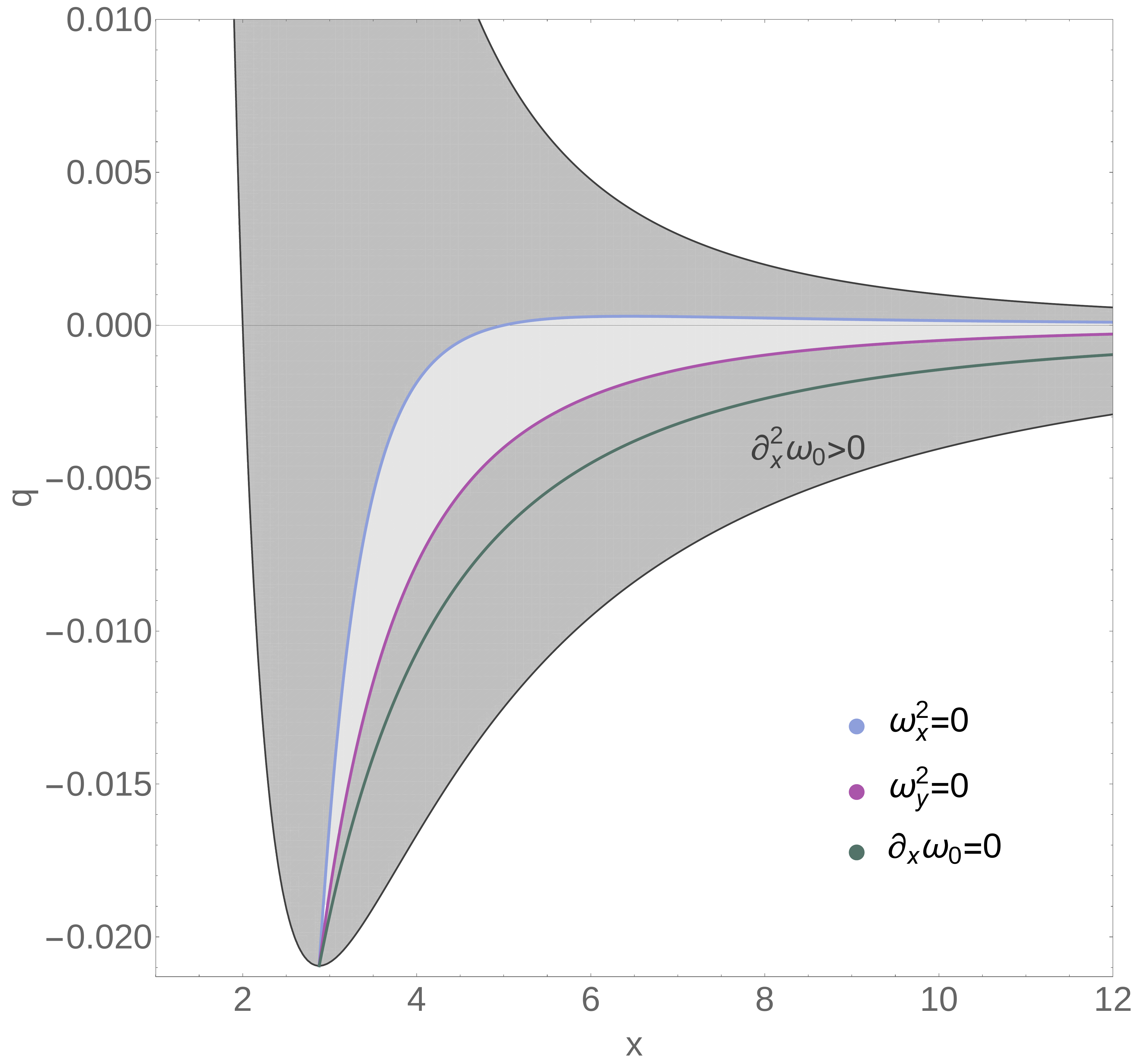} \\[1mm]
           \hspace{1.2cm}  $c)$
        \end{tabular}}
 \caption{\label{fig:frequency_max}\small Behavior of the radial epicyclic and the orbital frequencies. The curves $\partial_x\omega_x =0$ and $\partial^2_x\omega_x =0$ are represented in green and orange, respectively, in the plots a) and b). The function $\partial^2_x\omega_x$ is positive in the region right to the orange curve. The two curves intersect at the inflection point of the function $\omega_x(x)$, which occurs at $q_{crit}\approx -0.00034$. In plot c) we illustrate the curve $\partial_x\omega_0 =0$, while the function $\partial^2_x\omega_0$ is positive in the whole domain of existence of the timelike circular orbits. In our usual notations the domain of existence is depicted in dark gray, while the region of stability of the circular orbits is represented in light gray. }
\end{figure}

The analysis of the possible extrema of the epicyclic frequencies is presented in fig. $\ref{fig:frequency_max}$ where we plot the curves $\partial_x\omega_x =0$ and $\partial^2_x\omega_x =0$, and $\partial_x\omega_0 =0$, respectively. The critical value of the quadrupole moment $q_{crit}\approx -0.00034$, which separates the two regions of qualitatively different behavior of $\omega_x$ for negative quadrupole moments, corresponds to the minimum of the curve $\partial_x\omega_x =0$, or the inflection point of the function $\omega_x(x)$. In the figures it coincides with the intersection point of the curves $\partial_x\omega_x =0$ and $\partial^2_x\omega_x =0$. The curve $\partial^2_x\omega_0 =0$ does not intersect the domain of existence of the circular orbits, and $\partial^2_x\omega_0$ is always positive there. The positions of the minima of $\omega_x$ and $\omega_0$ tend to infinity when $q\rightarrow 0$. Examples of the three qualitatively different types of behavior of the characteristic frequencies are illustrated in fig. $\ref{fig:frequency}$ for fixed values of the quadrupole moment belonging to the relevant ranges.

\section{Non-linear resonances}

In the linear approximation we obtained two oscillator equations for the radial and vertical perturbations of the circular orbits with decoupled frequencies. However, if we consider higher order corrections, non-linear terms will be also introduced including coupling between the epicyclic frequencies. This coupling gives rise to different resonances, which are typically excited in non-linear systems. The exact form of the non-linear equations depends on the physical processes in the accretion disk, which are not well-understood. However, irrespective of their exact excitation mechanism, the formation of resonances is a natural manifestation of the non-linear effects. The analytical and numerical studies of some simple models of accretion confirm that they are indeed present, and  play an important role in the dynamics of the system.

One of the simplest situations when resonances can be exited is if we include a $\xi^x\xi^y$ term in the equation for the vertical oscillations. Then, it takes the form

\begin{eqnarray}
\frac{d^2 \xi^y}{dt^2} + \omega_y^2 \xi^y = - \omega_y^2 h \cos(\omega_x t) \xi^y ,
\end{eqnarray}
where $h$ is a constant. We obtain the Mathieu equation, which is known to describe parametric resonances for ratios of the frequencies

\begin{eqnarray}
\frac{\omega_x}{\omega_y}= \frac{2}{n},
\end{eqnarray}
where $n$ is a positive integer. The smallest possible value of $n$ leads to the strongest resonance. Despite the simplicity of this resonance model, it was found to be an intrinsic property of thin, nearly Keplerian disks \cite{Abramowicz:2003}, \cite{Rebusco}, \cite{Horak}.

Another possible source of resonant phenomena is the excitation of forced resonances. For example, the equation for the vertical oscillations may represent a forced non-linear oscillator  with a force frequency equal to the radial  epicyclic frequency

\begin{eqnarray}
\frac{d^2 \xi^y}{dt^2} + \omega_y^2 \xi^y
+ [{\rm non \; linear \; terms \; in} \: \xi^y] &=&
h(r) \cos(\omega_x t).
\end{eqnarray}

Resonances occur for integer ratios between the frequencies $\omega_y = n\omega_x$, and due to the non-linear terms linear combinations of the frequencies can be also present in the resonant solution. A third possibility is to obtain resonances due to a coupling between one of the epicyclic frequencies and the orbital frequency. Unlike the previous cases of coupling between the two epicyclic frequencies, which can be realized in numerous physical situations, such Keplerian resonances are less motivated. However, there is no physical reason restricting their existence.

The described resonance phenomena can give an explanation for the twin-peak oscillations in the accretion disk, provided that the ratio of the coupled frequencies is chosen appropriately, and the lower and the upper observable frequencies $\nu_L$ and $\nu_U$ are suitably identified. For the Schwarzschild black hole the observed ratio $\nu_U: \nu_L = 3:2$ can be reproduced in the three resonant models in the following ways.  Since it is always satisfied that $\omega_y > \omega_x$, the lowest possible parametric resonance is for $n=3$, i.e. we have $2\omega_y = 3\omega_x$. Making the identification $\nu_U = \nu_y= \omega_y/2\pi$ and  $\nu_L=\nu_x= \omega_x/2\pi$ we achieve the required $3:2$ ratio. Considering the forced resonances, it is impossible to choose a value of $n$ so that $\omega_y:\omega_x = 3:2$. However, we can identify some of the observable frequencies with a linear combination of the epicyclic frequencies. Choosing $\nu_U = \nu_y + \nu_x= (\omega_y +\omega_x)/2\pi$ and  $\nu_L=\nu_y$ we can reproduce the $3:2$ ratio by means of the $n=2$ forced resonance, or $\omega_y:\omega_x = 2:1$. Another possibility is the identification $\nu_U = \nu_y$ and  $\nu_L=\nu_y -\nu_x= (\omega_y -\omega_x)/2\pi$, where the $n=3$ forced resonance ensures the observed ratio, i.e. $\omega_y:\omega_x = 3:1$. For the Schwarzschild solution the simplest cases for Keplerian resonances, such as $\omega_0:\omega_x = 3:2$, $\omega_0:\omega_x = 2:1$, or $\omega_0:\omega_x = 3:1$, do not provide new possibilities for modelling the twin-peak oscillations since the orbital frequency coincides with the vertical epicyclic frequency.

\begin{figure}[t!]
    		\setlength{\tabcolsep}{ 0 pt }{\small\tt
		\begin{tabular}{ cc}
           \includegraphics[width=0.5\textwidth]{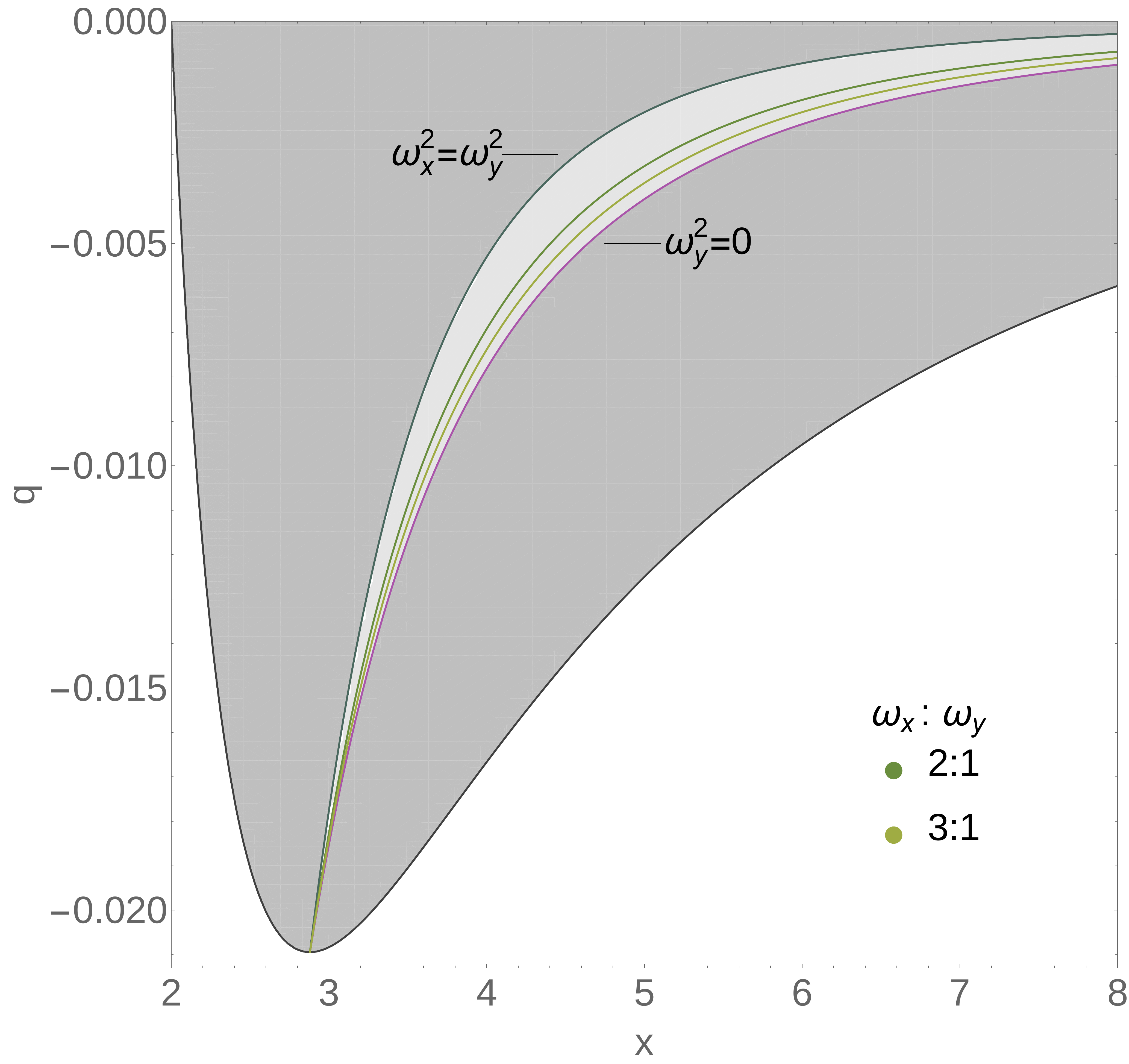}
		   \includegraphics[width=0.5\textwidth]{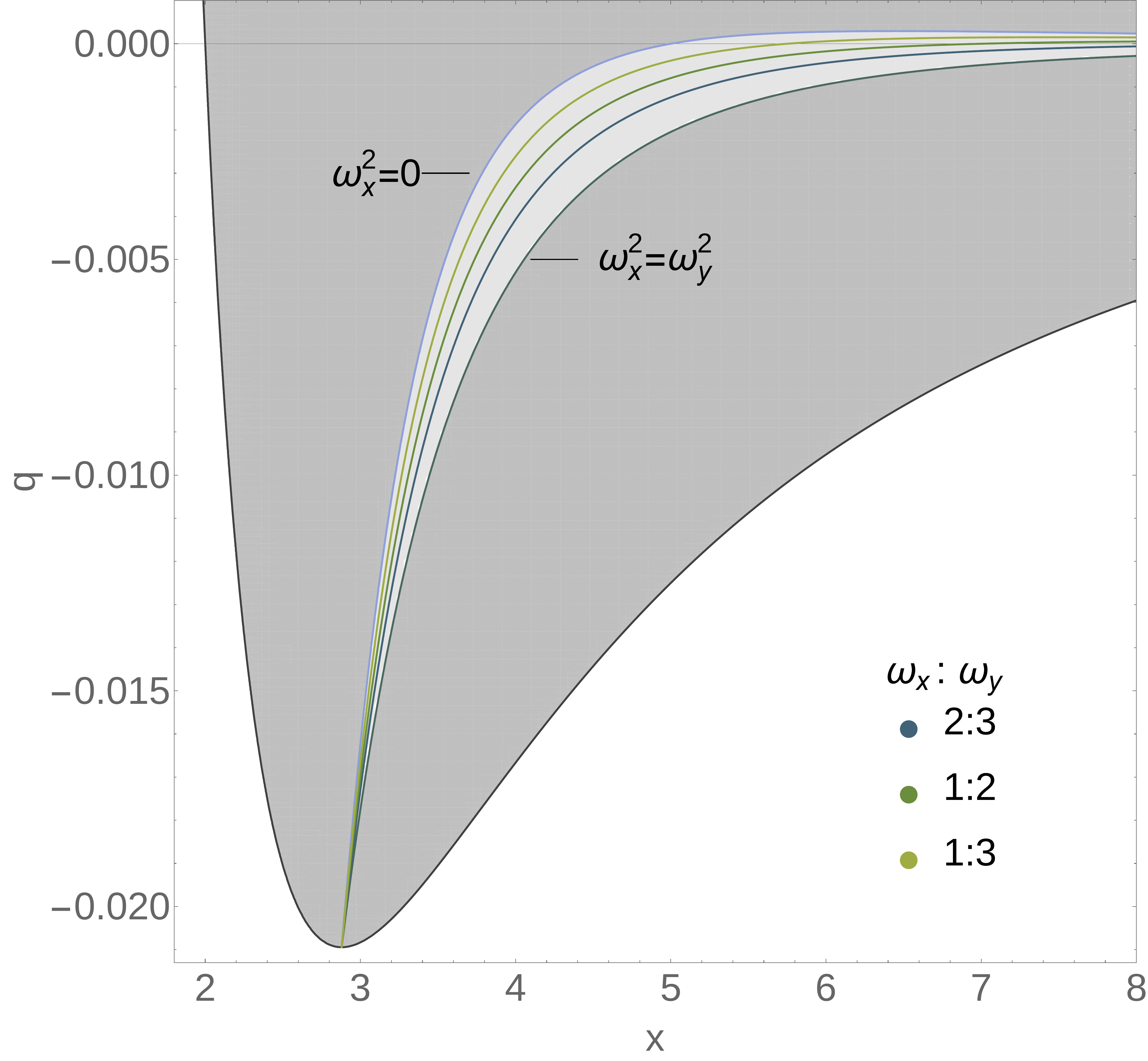} \\[1mm]
           \hspace{1.2cm}  $a)$ \hspace{7cm}  $b)$ \\[3mm]
           \includegraphics[width=0.5\textwidth]{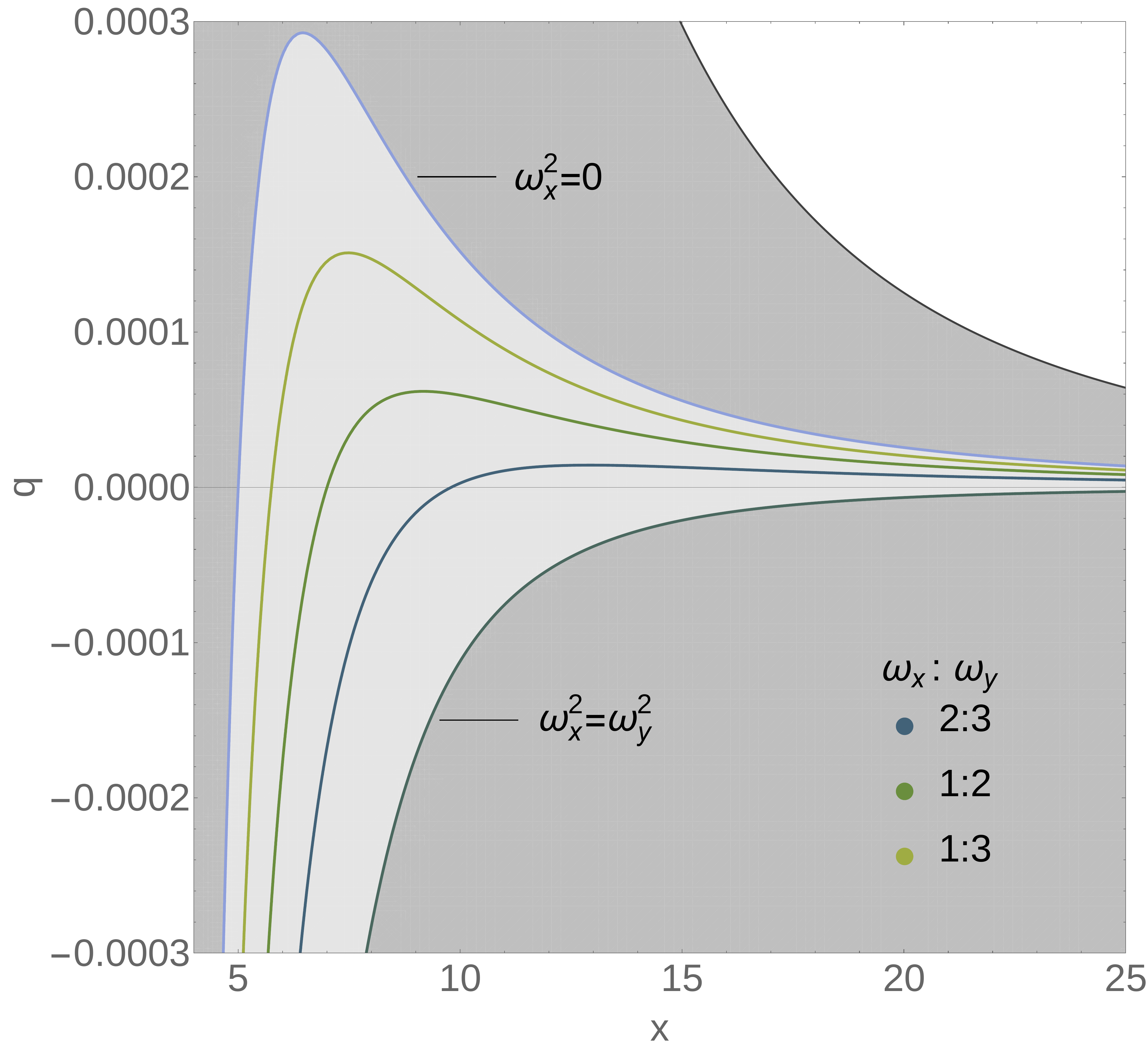} \\[1mm]
           \hspace{1.2cm}  $c)$
        \end{tabular}}
 \caption{\label{fig:res1}\small Location of the parametric and forced resonances for ordering between the epicyclic frequencies a) $\omega_x > \omega_y$, and b) $\omega_x <\omega_y$. In the zoomed version of the case $\omega_x <\omega_y$ c) we can see the location for the  Schwarzschild black hole in the limit $q=0$. }
\end{figure}

\begin{figure}[t!]
    		\setlength{\tabcolsep}{ 0 pt }{\small\tt
		\begin{tabular}{ cc}
           \includegraphics[width=0.5\textwidth]{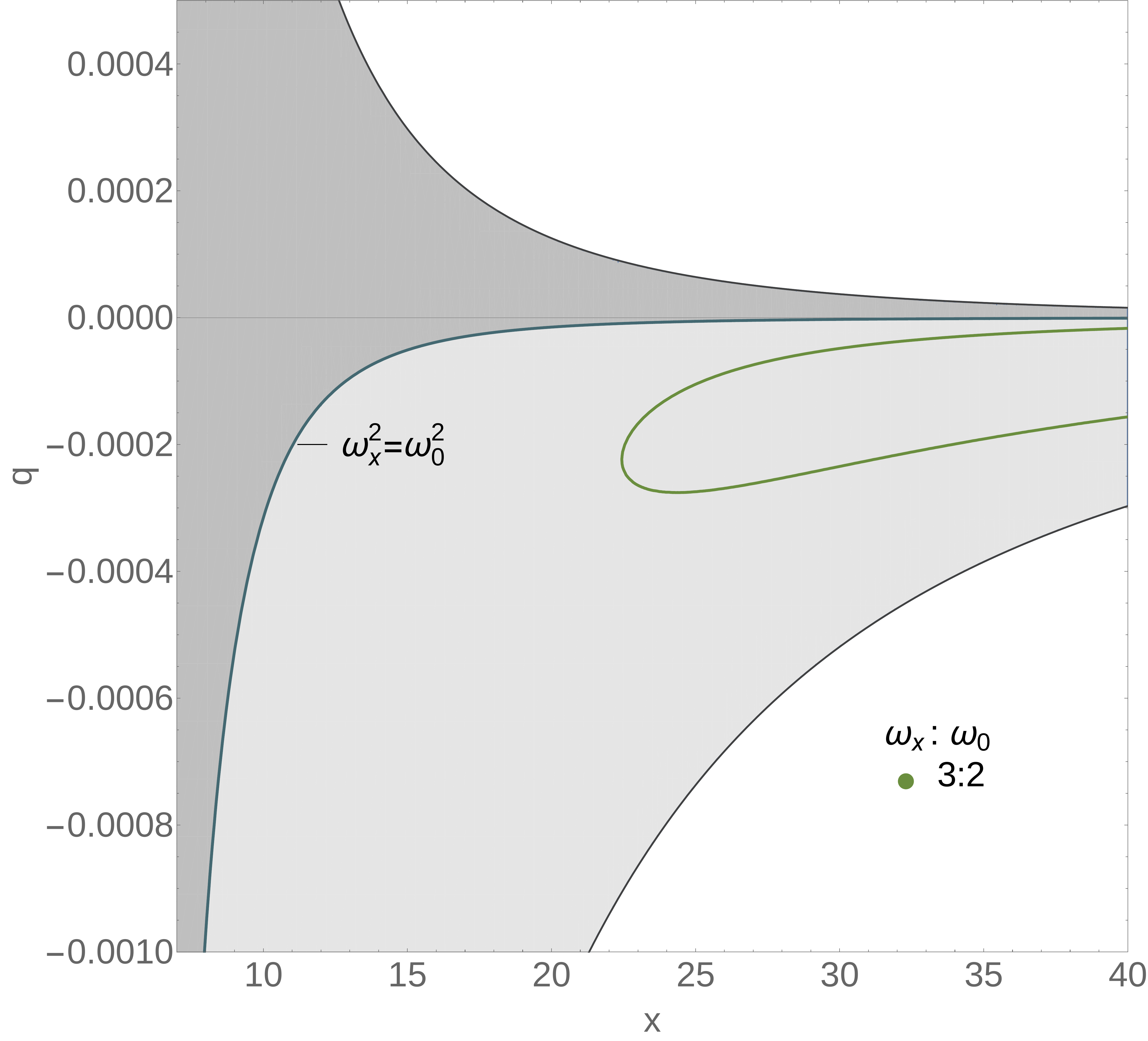}
		   \includegraphics[width=0.5\textwidth]{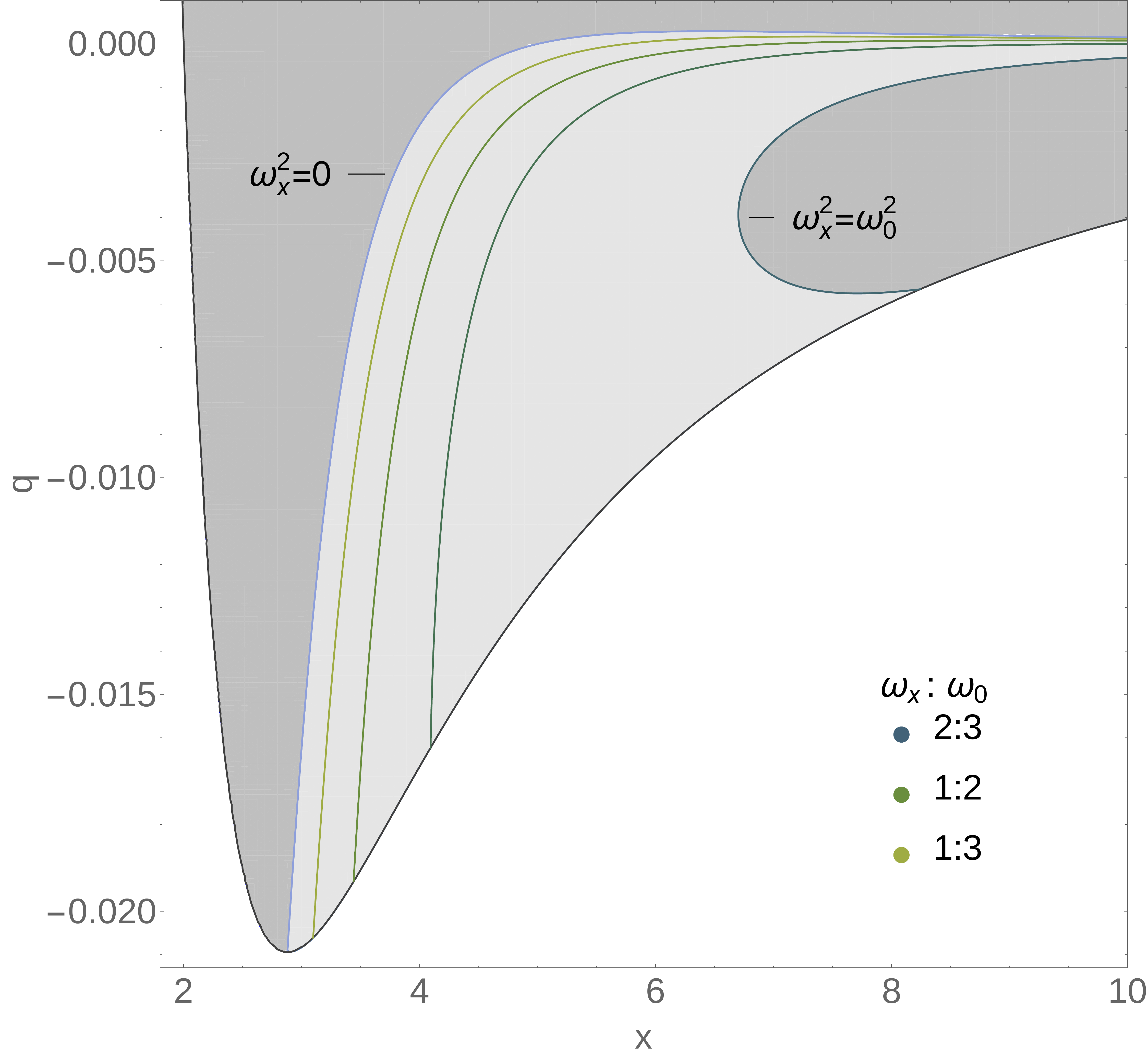} \\[1mm]
           \hspace{1.2cm}  $a)$ \hspace{7cm}  $b)$ \\[3mm]
           \includegraphics[width=0.5\textwidth]{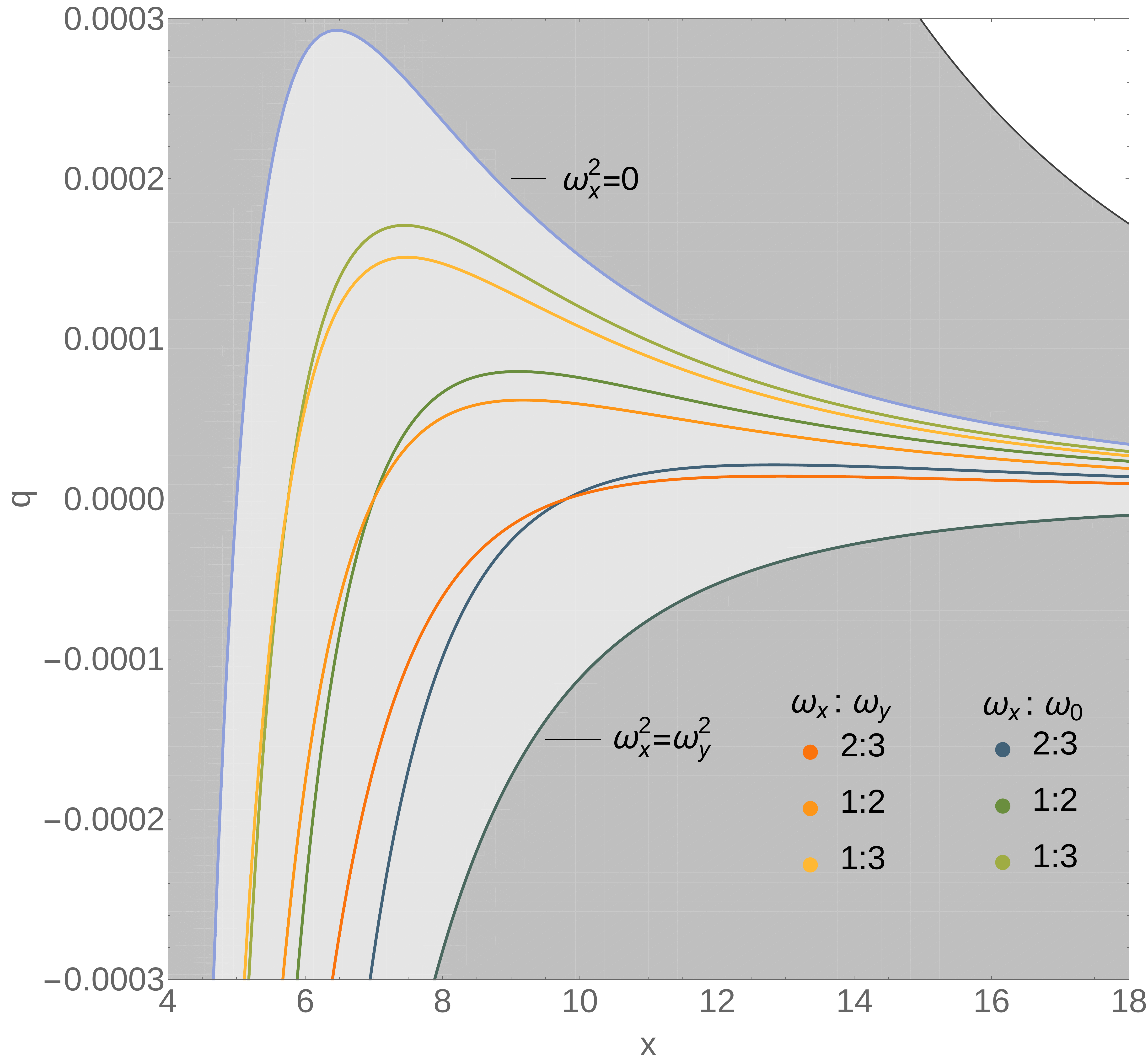} \\[1mm]
           \hspace{1.2cm}  $c)$
        \end{tabular}}
 \caption{\label{fig:res2}\small Location of the Keplerian resonances with coupling between the radial epicyclic frequency and the orbital frequency. The case $\omega_x >\omega_0$ is represented in a), while b) and c) illustrate the ordering $\omega_x <\omega_0$. For $\omega_x >\omega_0$ the location of the possible $2:1$ and $3:1$ resonances has no cross-section with the domain of existence of the circular orbits. In the zoomed image c) we compare the locations of the Keplerian resonances with the parametric and forced ones for $\omega_x <\omega_y$.  }
\end{figure}

\begin{figure}[t!]
    		\setlength{\tabcolsep}{ 0 pt }{\small\tt
		\begin{tabular}{ cc}
           \includegraphics[width=0.5\textwidth]{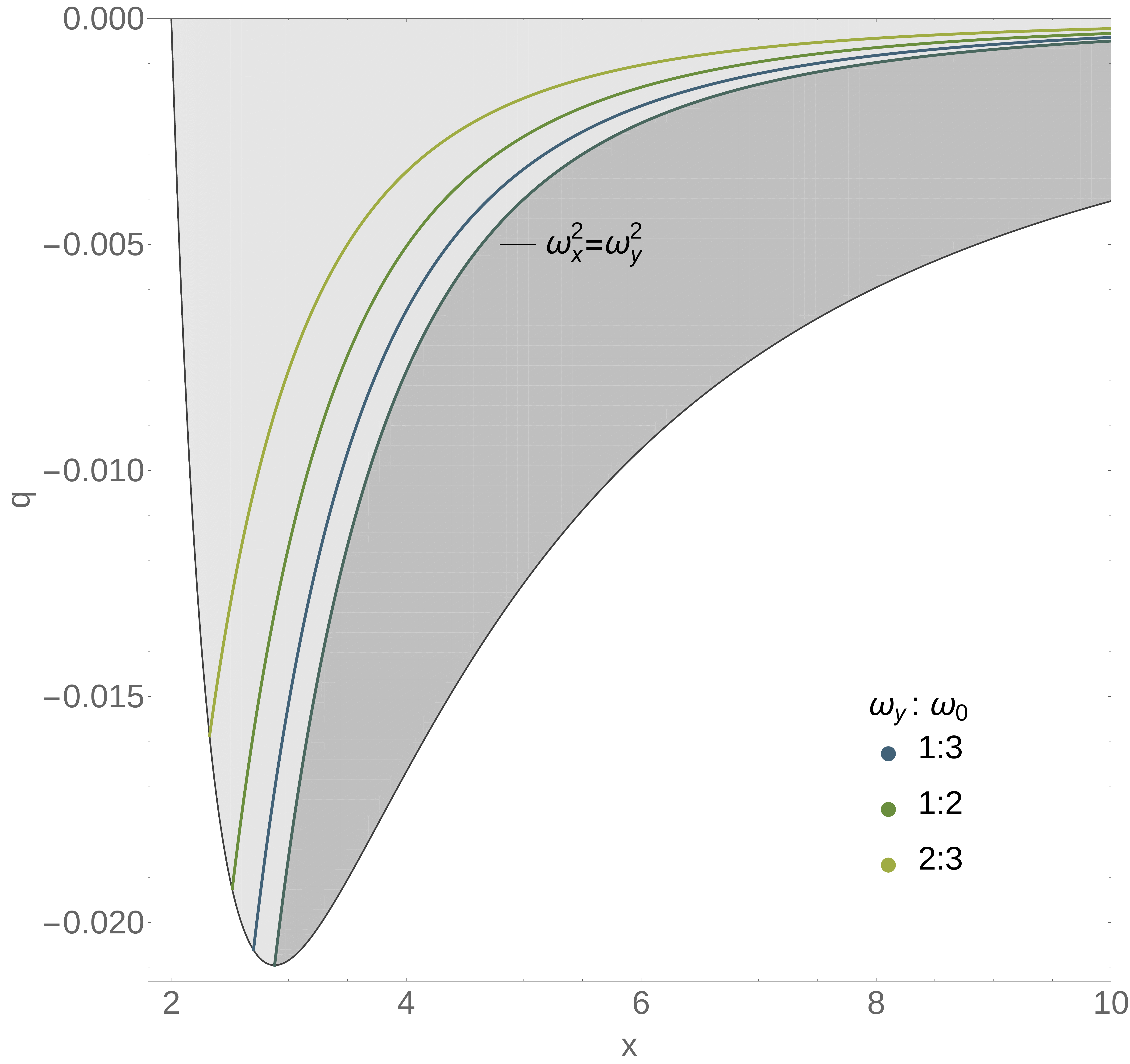}
		   \includegraphics[width=0.5\textwidth]{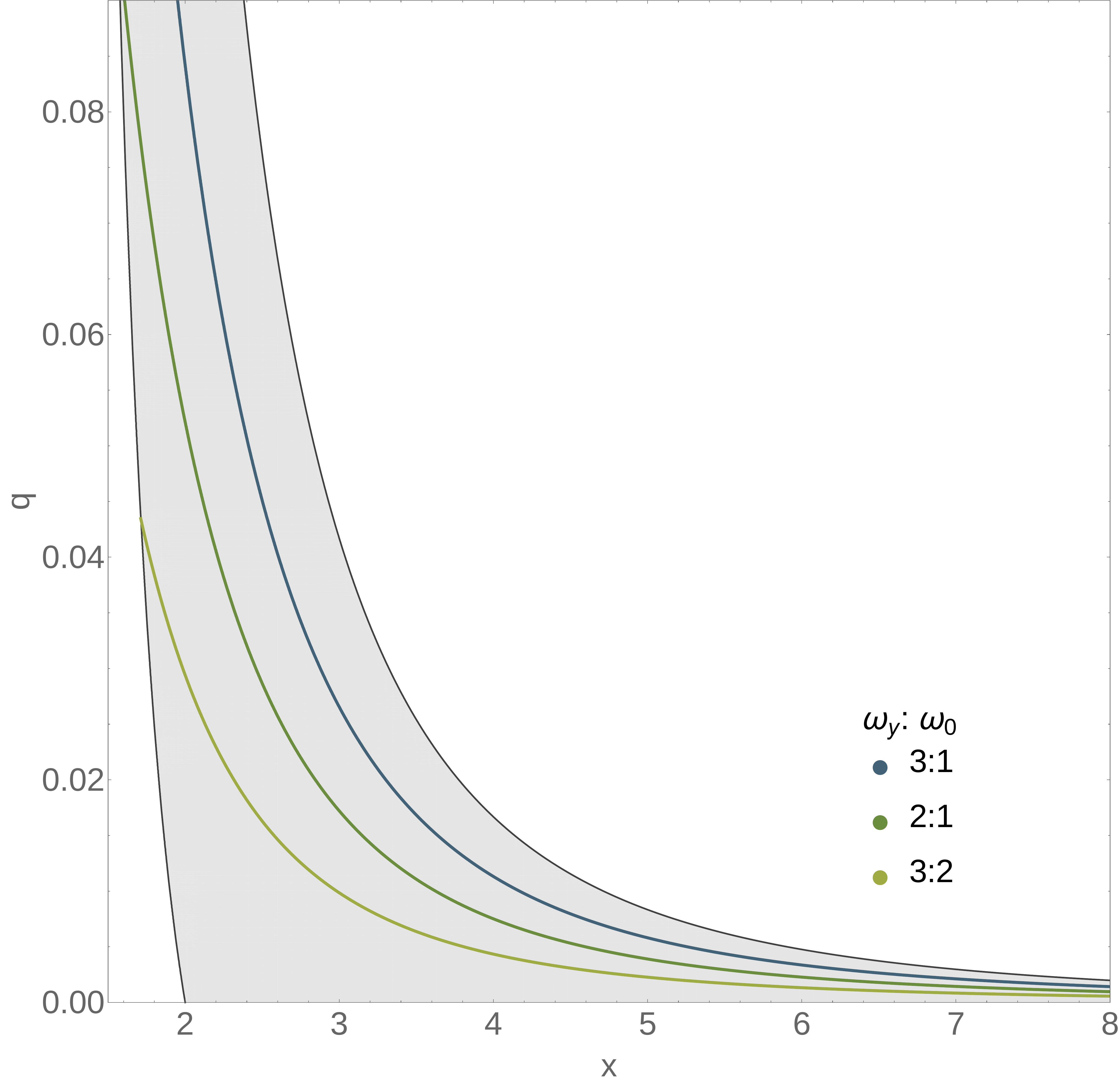} \\[1mm]
           \hspace{1.2cm}  $a)$ \hspace{7cm}  $b)$
        \end{tabular}}
 \caption{\label{fig:res3}\small Location of the Keplerian resonances with coupling between the vertical epicyclic frequency and the orbital frequency for a) $\omega_y<\omega_0$ and b) $\omega_y>\omega_0$.}
\end{figure}

In the case of the distorted Schwarzschild solution the diversity of the possible resonances increases, since we have regions in the parametric space where the ordering of the epicyclic frequencies changes to $\omega_y \leq\omega_x$. Moreover, the degeneracy between the orbital and vertical epicyclic frequencies is broken, and Keplerian resonances can exist independently. In addition to the types of resonances available for the Schwarzschild solution, the distorted black hole allows for parametric resonances with $n=1$ and $n=2$ in the regions where $\omega_y \leq\omega_x$ is satisfied, which are supposed to be stronger than the $n=3$ one (see e.g. \cite{Abramowicz:2003}). For $n=1$ the observed $3:2$ ratio can be achieved by identifying the lower and the upper frequencies as $\nu_U = \nu_y + \nu_x$ and  $\nu_L=\nu_x$, while for $n=2$ we have $\nu_U = 3\nu_y =3\nu_x$ and $\nu_L=2\nu_x=2\nu_y$. The new possibilities for forced resonances include ratios of the epicyclic frequencies like $\omega_y:\omega_x = 1:2$ and $\omega_y:\omega_x = 1:3$, which result in the observable frequencies $\nu_U = \nu_y + \nu_x$,  $\nu_L=\nu_x$, and $\nu_U=\nu_x$, $\nu_L = \nu_x - \nu_y$, respectively. The Keplerian resonances can be excited independently with combinations like  $\omega_0:\omega_x = 3:2$ ($\nu_U = \nu_0$, $\nu_L = \nu_x$) , $\omega_0:\omega_x = 2:1$ ($\nu_U = 3\nu_x$, $\nu_L = \nu_0$), or $\omega_0:\omega_x = 3:1$ ($\nu_U = \nu_0$, $\nu_L = 2\nu_x$) in the regions where the ordering $\omega_0>\omega_x$ is valid, and with switched positions of the two frequencies $\omega_0\leftrightarrow\omega_x$ where the inequality $\omega_0<\omega_x$ is satisfied. The location of the described resonances as a function of the quadrupole moment is illustrated in figs. $\ref{fig:res1}$-$\ref{fig:res3}$.

\section{Conclusion}

Astrophysical black holes are not expected to be isolated, but interacting with external gravitational sources like binary companions and accretion disks. If taken into account, the influence of the external matter can lead to significant observational effects, especially when particle and light propagation in the black hole spacetime is considered. In this paper we study how the impact of the external matter can modify the interpretation of the quasi-periodic oscillations from the accretion disk by means of the resonance models. We consider in particular a distorted Schwarzschild black hole interacting with an external matter distribution possessing only a quadrupole moment. As a first step we examine the linear stability of the circular orbits in the equatorial plane and obtain analytical expressions for the epicyclic frequencies. For positive quadrupole moments the stability of the circular orbits is determined only by their stability with respect to perturbations in radial direction, similar to the isolated Schwarzschild solution. However, there exists an upper limit for the quadrupole moment, for which stable orbits can exist. For negative quadrupole moments the vertical epicyclic frequency can become negative, and introduce an instability with respect to perturbations in vertical direction. Therefore, the stability of the circular orbits is determined by the condition that both epicyclic frequencies are simultaneously positive. For a fixed quadrupole moment this restricts the radial positions of the stable orbits to a finite interval, so they belong to a finite annulus, rather than extending to infinity like for the isolated Schwarzschild solution. In addition, there is a lower limit of the quadrupole moment, for which stable orbits are possible. These properties can be used to determine the region of validity of the distorted Schwarzschild solution by physical reasons. Depending on the desired structure of the equatorial circular orbits for the global solution, the boundary of the region of validity should be chosen either within or outside the region of stability for the distorted Schwarzschild solution.

We further examine the properties of the epicyclic frequencies, which show some important qualitative differences from the isolated Schwarzschild case. The vertical epicyclic and the orbital frequencies do not coincide anymore. It addition, it is not always satisfied that the vertical epicyclic frequency is larger than the radial one. There are regions in the parametric space where the opposite inequality is true. In the same way there can be different ordering between the vertical or radial epicyclic frequency and the orbital frequency. All these possibilities enable the excitation of much more diverse types of resonances compared to the isolated Schwarzschild solution, resulting  from different physical processes in the accretion disk. They can be stronger than in the Schwarzschild case, and located at very different radial distances, depending on the value of the quadrupole moment.

\section*{Acknowledgments}
We gratefully acknowledge support by the DFG Research Training Group 1620 ``Models of Gravity''  and the COST Actions CA16214 and CA16104.  P.N. is partially supported by the Bulgarian NSF Grants DM 18/3 and  KP-06-H38/2.

\section*{Appendix: Christoffel symbols}

We present the non-zero connection coefficients for the distorted Schwarzschild solution:

\begin{eqnarray}
\Gamma^{t}_{tx}&=&\frac {1}{( x^2 -1)} +{\cal{U}}_{,x}\, , \nonumber \\
\Gamma^{t}_{ty}&=&{\cal{U}}_{,y}\, , \nonumber \\
\Gamma^{x}_{tt}&=&\frac {( x-1)}{( x+1) ^{3}}\left[1+ ({x}^{2}-1)\,{\cal{U}}_{,x}\right]e^{4{\cal{U}}-2V}\,,\nonumber\\
\Gamma^{x}_{xx}&=&-\frac{1}{x^{2}-1}+V_{,x}-{\cal{U}}_{,x}\,,\nonumber\\
\Gamma^{x}_{xy}&=&V_{,y}-{\cal{U}}_{,y}\, ,\nonumber \\
\Gamma^{x}_{yy}&=&-\frac {( x-1)}{(1-y^{2})}\left[1+(x+1)(V_{,x}-{\cal{U}}_{,x})\right]\,,\nonumber\\
\Gamma^{x}_{\phi\phi}&=&-(x-1)(1-y^{2})\left[1-(x+1)\,{\cal{U}}_{,x}\right]e^{-2V}\,, \nonumber \\
\Gamma^{y}_{tt}&=&\frac{( x-1)}{(x+1) ^{3}}(1-{y}^{2})\,{\cal{U}}_{,y}e^{4{\cal U}-2V}\,,\nonumber\\
\Gamma^{y}_{xx}&=&\frac{(1-y^{2})}{(x^{2}-1)}\left({\cal{U}}_{,y} -V_{,y}\right)\,,\nonumber\\
\Gamma^{y}_{xy}&=&\frac{1}{x+1}+V_{,x}-{\cal{U}}_{,x}\,,\nonumber\\
\Gamma^{y}_{yy}&=&\frac{y}{1-y^{2}}+V_{,y}-{\cal{U}}_{,y}\,,\nonumber\\
\Gamma^{y}_{\phi\phi}&=&(1-y^{2})\left[y+(1-y^{2})\,{\cal{U}}_{,y}\right]e^{-2V}\,, \nonumber \\
\Gamma^{\phi}_{x\phi}&=&\frac {1}{( x+1)}-{\cal{U}}_{,x}\, \nonumber \\
\Gamma^{\phi}_{y\phi}&=&-\frac {y}{(1-y^2)} -{\cal{U}}_{,y}\, .
\end{eqnarray}

\end{document}